\documentclass[10pt,leqno]{amsart}
\usepackage{graphicx}
\usepackage{indentfirst,csquotes}

\usepackage[english]{babel}
\usepackage[T1]{fontenc}
\usepackage[utf8]{inputenc}
\usepackage{amsmath}
\usepackage{amssymb}                                          
\usepackage{graphicx}

\usepackage{natbib}

\usepackage{mathtext}
\usepackage[hyperfootnotes=false]{hyperref}
\usepackage{color}
\usepackage{epsf}
\usepackage{amsfonts}
\usepackage{gensymb}
\usepackage{amssymb}
\usepackage{float}
\usepackage{ulem}
\usepackage{diagbox}

\def\tightleading{1.1}
\def\doubleleading{1.6}
\def\baselinestretch{\doubleleading}

\let\tightenlines=\tighten

\tightenlines

\def\ssr{Space Sci. Rev.}

\def\apj{ApJ}
\def\apjl{ApJL}
\def\apjs{ApJ Suppl. Ser.}
\def\apss{Astroph.Sp.Sci.}
\def\aap{A\&A}

\def\araa{ARAA}

\def\prl{Phys. Rev. L}

\def\pasj{PASJ}

\def\solphys{Solar Physics}

\usepackage{amssymb,amsthm,amsmath}
\usepackage{xcolor,paralist,hyperref,titlesec,fancyhdr,etoolbox}

\titleformat{\section}[display]{\normalfont\huge\bfseries\centering}{\centering\chaptertitlename\thechapter}{10pt}{\Large}
\titlespacing*{\section}{0pt}{0ex}{0ex}

\hypersetup{ colorlinks=true, linkcolor=black, filecolor=black, urlcolor=black }
\def\proof{\noindent {\it Proof. $\, $}}

\usepackage{lipsum}

\begin{document}
\title{Multiwavelength Observations of the Precursors Before the Eruptive X4.9 Limb Solar Flare on 25 February, 2014: Pre-flare Current Sheet, Build-up of Eruptive Filament, Flare and Eruption Onset in the Frame of the Tether-Cutting Magnetic Reconnection Scenario} 

\author{I.\,N.~Sharykin$^{1*}$, I.\,V.~Zimovets$^{1}$, G.G.~Motorina$^{1,2}$, N.\,S.~Meshalkina $^{3}$ \\
}

\date{\today}
\address{Profsoyusnaya street, 84/32, Moscow, Russia}
\email{ivan.sharykin@phystech.edu}
\maketitle

\let\thefootnote\relax
\footnotetext{$^*$ {\it ivan.sharykin@phystech.edu} \\ $^1$ Space Research Institute of the Russian Academy of Sciences (IKI), Moscow 117997, Russia \\
$^2$ The Central Astronomical Observatory of the Russian Academy of Sciences at Pulkovo, Saint-Petersburg 196140, Russia\\
$^3$ Institute of Solar-Terrestrial Physics of Siberean Branch of Russian Academy of Sciences, Irkutsk 664033, Russia\\} 

\begin{abstract}
We present multiwavelength analysis of the pre-flare phase, as well as the onset of the powerful X4.9 near-limb eruptive solar flare on February 25, 2014 (SOL2014-02-25T00:39), revealing the tether-cutting geometry. This event provides an excellent opportunity to investigate pre-flare and flare energy release in details utilizing available large volume of observational data in different spectral ranges, suitable limb position of the flare, high power of its energy release, favorable spatial distribution of pre-flare and flare emission sources and well-observed eruption. We aim at determining relationship between the region of pre-flare energy release with the regions where the flare started to develop, and to investigate a detailed chronology of energy release during the pre-flare time interval and the beginning of the impulse phase. Using X-ray, ultraviolet (UV) and radio microwave data we found that the pre-flare energy release site was compact and localized in the vicinity of interaction (tether-cutting type) of larger-scale magnetic structures near the polarity inversion line of the magnetic field. The analysis indicates that a pre-flare current sheet could be in this region. Good correspondence between the location of the pre-flare and flare emission sources visible at the very beginning of the impulsive phase is shown. We found relationship between dynamics of the energy release in the pre-flare current sheet and formation of the future flare eruptive structure. The growth of the magnetic flux rope was associated with activation of plasma emissions, flows and an increase of UV radiation fluxes from the region where the pre-flare current sheet was located. The eruptive flux rope gradually grew due to ``feeding'' by magnetized plasma ejected from the reconnecting pre-flare current sheet. Finally, it is shown that the most probable trigger of the eruption was a local fast microflare-like magnetic reconnection in the pre-flare current sheet. Some local instability in the pre-flare sheet could lead to a transition from the slow to fast reconnection regime. As a result, an ejection from the sheet was initiated and the eruptive flux rope lost its stability. Then, the eruptive flux rope itself initiated formation of the main reconnecting flare current sheet as in the Standard Flare Model (under the flux rope) during its movement, and intense emissions associated with the impulsive phase were observed.  

\end{abstract} 

\bigskip 
{\bf Introduction} \\

The search for precursors of various large-scale natural phenomena affecting our life is necessary for their prediction and, accordingly, detection of the sources of future main energy release and its consequences. One of the examples is the search for precursors of earthquakes and tsunamis. Similar problems are considered in the Solar physics in application to the most large-scale phenomena of solar activity affecting space weather. We mean a search for precursors of the solar flares and coronal mass ejections (CMEs) in order to localize their future initial energy release and predict when it will be triggered. Study of the pre-flare activity has also fundamental interest for understanding physical processes typical for an active region (AR) when it prepares to future main flare energy release. From the practical point of view it is especially important to understand what types of precursors are typical for the most geoeffective solar flares, accompanied by energetic particle increases and powerful CMEs.

It is clear that the pre-flare energy release is associated with the dynamics of the magnetic field in the solar atmosphere and it can be detected in different ranges of electromagnetic spectrum. Therefore, the observational peculiarities of the precursors should be determined by the magnetic field topology and the mechanisms responsible for the formation of unstable magnetic field configurations. For example, according to the standard model (SM) of an eruptive solar flare \citep[e.g.][]{Hirayama1974,Magara1996,Tsuneta1997} the beginning of energy release is associated with loss of stability of a twisted magnetic flux rope extended along a magnetic polarity inversion line (PIL). Thus, in the framework of SM, precursors should show the development of magnetic flux rope instability at the phase of slow rise before eruption. Other flare models will assume their own manifestations of pre-flare energy release. Here we do not discuss in details all possible scenarious as we will focus on particular model explaining formation of the magnetic flux rope and the flare onset.

One of the most common phenomena in ARs producing solar flares is appearance of sigmoidal structures and magnetic flux ropes along the PIL \citep[e.g., reviews of ][]{Gibson2006, Toriumi2019} revealing non-potentiality of the magnetic field and excess of free magnetic energy due to strong electric currents in an AR. These configurations can be unstable due to resistive mechanisms or ideal MHD instabilities. One of the most popular scenarios is called the tether-cutting mechanism \citep[][]{Moore2001} assuming magnetic reconnection (tether-cutting magnetic reconnection, TCMR) within magnetic flux ropes initiated between individual interacting crossed and sheared magnetic loops. As a result of TCMR, a current sheet is formed in the region of a magnetic field tangential discontinuity (where we have a non-zero magnetic field). TCMR produces magnetic field lines elongated along the PIL, which can interact with an overlying magnetic arcade. If reconnection process is rather slow, TCMR results in a gradual formation of a twisted magnetic flux rope, which at some point can become eruptive and trigger a solar flare within the framework of the SM. If reconnection is explosive, then flare is initiated immediately without a long phase of magnetic flux rope formation \citep[e.g., ][]{Liu2013, Sharykin2018}. In these two cases, the precursors should be associated with the interaction of two crossed magnetic loop systems carrying strong electric currents and the formation of a CS in the region of their contact region close to the PIL. The destabilization of the pre-flare CS will result in the flare onset observing as either the instability of the magnetic flux rope or an explosive release of magnetic energy in the form of plasma heating and particle acceleration without the development of eruption (case of a ``closed'' flare or confined eruption).  

The possible observational manifestations of the pre-flare energy release in the TCMR frame can be various in different parts of electromagnetic spectrum and show different physical processes: strong plasma heating, appearance of nonthermal particles, plasma flows along magnetic field, plasma turbulence, magnetic field restructuring, eruptions of different scales, etc. \cite{Chifor2006, Chifor2007} found that the pre-flare SXR emission sources are likely due to magnetic reconnection occurred at the PIL where crossed magnetic loops interacted and eruption was triggered. \cite{Woods2017} concluded that one of the most probable reasons for the precursor activity in the form of transient brightnings and high speed plasma flows at the PIL for the studied flare was internal reconnection in the flux rope in the frame of the TCMR. Another pre-flare activity in the form of hot channel elongated along the PIL was shown in the pre-flare phase (one hour before the onset) for a closed M1.2 class flare in the article of \cite{Sharykin2018, Sharykin2020}. According to the results of nonlinear force-free extrapolation of the magnetic field, it was shown that the hot channel is associated with a rather low (<4 Mm) magnetic structure with a very large shear. It was discussed that strong heating in the channel can occur as a result of slow pre-flare magnetic reconnection within the tether-cutting geometry. Estimation of the magnetic field in the region of the pre-flare energy release site was made in \cite{Wang2017} {by the} analysis of microwave 1-18~GHz emission observed by eOVSA (expanded Owen Valley Solar Array). It was shown that the maximum value of the magnetic field (500-1000~G) was at the beginning of the pre-flare burst. This fact and observations in the optical and UV ranges mean that the precursors were associated with the low-lying region above the PIL, where the sheared magnetic loops were observed. 

Another well-known pre-flare phenomenon is the neutral line sources (NLS, see e.g. \cite{Uralov2006, Uralov2008, AbramovMaximov2015a, Bakunina2015}) seen in the microwave range. These sources are observed between sunspots and are characterized by high brightness temperatures from several MK to $\sim10$~MK, the spectral maximum in the range of 5-7~GHz, sharply decreasing spectrum and the life time as high as of several days. One of the most probable explanations for the features of NLS radiation is the thermal gyroresonance radiation of a hot plasma (up to 10 MK) in a strong magnetic field ($\sim$1000 G). The most popular scenario of the appearance of NLS involves the presence of a twisted magnetic flux rope at the PIL interacting with an external magnetic field enveloping it. However one can assume internal TCMR-like energy release inside the magnetic flux rope. \cite{Kudriavtseva2021} showed that the NLS is associated with the sheared magnetic flux rope.

Episodic reconnection bursts in the sigmoidal structures can be observed as oscillations at the PIL where TCMR situation is possible. Such observations were analyzed by \cite{Zhou2016} using UV spectroscopy and images. It was concluded that non-stationary pre-flare energy release was associated with the standing wave of the fast magnetoacoustic kink mode stimulated by reconnection episodes at the PIL. 

All these mentioned works show that the tether-cutting scenario has a lot of observational manifistations during the pre-flare phase. This scenario is quite natural for solar flare and should be extensively investigated. From our point of view, there were not so much detailed multiwavelength targeted studies of the TCMR processes. In particular, there were no attempts to localize the pre-flare and the flare tether-cutting current sheets like it was done for SM \citep[e.g., ][]{Su2013, Longcope2018}. It is also important to trace how an eruptive flux-rope is formed and how it is connected with the pre-flare TCMR current sheet. We also need more quantitative studies showing plasma energy release rate during the precursor phase.

The main aim of this paper is to perform a detailed multiwavelength study of the pre-flare energy release in the frame of the TCMR scenario. There are a few specific tasks and corresponding questions considered in this paper:
\begin{enumerate}
	\item To determine spatial connections between a pre-flare energy release site and a site where we detected the flare onset. Are there any signatures of the magnetic reconnection in the region of the pre-flare energy release site and can we find the signatures of the pre-flare current sheet?
	\item In the case of the verification of the existence of the pre-flare current sheet we need to determine characteristics of the thermal plasma and energy release rate.
	\item To determine the detailed chronology of the pre-flare energy release processes, preparation of the AR to the future flare, transition from the pre-flare state to the flare and eruption.
\end{enumerate}

The results obtained through this paper will allow us to advance in understanding of energy accumulation and release processes in flaring ARs, as well as improve our comprehension of the physics behind solar flares. In particular, it is expected that more deep focus on the pre-flare process helps us to find new valuable physical principles which could be used for predicting solar flares in future.

The paper (consisting of three sections) is organized as follows: flare selection criteria, data description, flare overview, observational results (ultraviolet, X-ray and microwave emission analysis), conclusions and discussion of the obtained results. \\


{\bf 1. Data and Flare Selection} \\


{\bf{\it 1.1. Event Selection and Used Observational Data}} \\

Along with the main requirement for a flare to be eruptive and follow TCMR morphology we assume a number of additional observational criteria of the event selection:
\begin{enumerate}
	\item The flare must be eruptive with well developed coronal mass ejection (CME) and clearly observed eruption dynamics in the low solar corona by the AIA without saturation. The onset of eruption and dynamics of the flare loops during initial energy release should be clearly observed. In this sense, a flare near the limb will be an ideal candidate;
	\item Sheared loops should be detected in the EUV range to find the TCMR-like magnetic field structure;
	\item The flare should be isolated from previous events. It means that there is no overlapping between a ``tail'' of a previous flare with the onset of the studied one;
	\item Pronounced signs of presence of accelerated electrons during the impulsive phase;
	\item The availability of the microwave and X-ray observations for the pre-flare, flare onset and the impulsive phase time periods. It is necessary for study of the plasma heating and electrons acceleration during all stages;
	\item The solar flare should be sufficiently strong (high M or X-class flare) to achieve good RHESSI count statistics for X-ray image reconstruction around the flare onset and even during the pre-flare time period;
	\item The flare must be accompanied by the photospheric energy release seen in HMI data. One can make analysis of the initial photospheric flare energy release with high temporal resolution using 1.8-seconds HMI level-1 filtergrams. Using such data we will be able to localize primary photospheric energy release and to compare with the pre-flare energy release site.
\end{enumerate}

To find solar flare with the confirmed TCMR geometry we decided to study previous papers where the tether-cutting model was discussed. In result we found such event matching criteria mentioned above in the paper of \cite{Chen2014}. The flare had GOES X4.9 class and occurred on February 25, 2014. According to the GOES flare catalog the flare start was around 00:39~UT and the peak was at 00:49~UT. The flare site was located in the NOAA AR~11990 with the heliographic coordinates S15E65 ($x=-849^{\prime\prime}$, $y=-199^{\prime\prime}$). In the work of \cite{Chen2014} pre-flare processes were briefly and qualitatively discussed. In our work we will carry out a detailed analysis of the pre-flare energy release transforming into the flare onset in accordance with the paper goals and questions stated in the Introduction. In particular, we want to localize the position of the TCMR current sheet and study pre-flare and the flare onset processes around it both in qualitative and quantitative way.

To solve research tasks we will use the following data. To determine spatial structure of the hottest plasma (above 10~MK) and distribution of the nonthermal electrons we will analyze X-ray images from the Reuven Ramaty High Energy Solar Spectroscopic Imager (RHESSI, \cite{Lin2002}). Additional quantitative information about hot plasma will be extracted from the GOES X-ray lightcurves. EUV level 1.5 images from the Atmospheric Imaging Assembly (AIA, \cite{Lemen2012}) onboard the Solar Dynamics Observatory (SDO, \cite{Pesnell2012}) will be used to obtain more detailed information about plasma heating in the wide temperature range (0.5-16~MK), spatial dynamics of magnetic structures and plasma flows in the AR. To determine magnetic changes in the flare region we will use vector magnetograms from the Helioseismic Magnetic Imager (HMI, \cite{Scherrer2012}). 

To investigate photospheric flare energy release we will use HMI level~1 data, called filtergrams. These images are made for six wavelength positions and different polarization modes by two HMI cameras. The used filtergrams processing methods are presented in the works \cite{Sharykin2017, Sharykin2020} and based on subtraction of the corresponding (certain polarization and wavelength channel) pre-flare filtergram from the flare one.

Additional information about energy release in the pre-flare region will be obtained from microwave 17 and 34~GHz images from the Nobeyama Radioheliograph (NoRH). To determine quantitative characteristics of the pre-flare plasma we will also analyze RHESSI X-ray spectra and two-channel (0.5-4 and 1-8~\AA{}) GOES lightcurves. More detailed comments on the used techniques will be described in the following corresponding sections. \\

{\bf{\it 1.2. Flare Overview}} \\

\begin{figure*}    
    \centering
	\centerline{\includegraphics[width=1.0\textwidth,clip=]{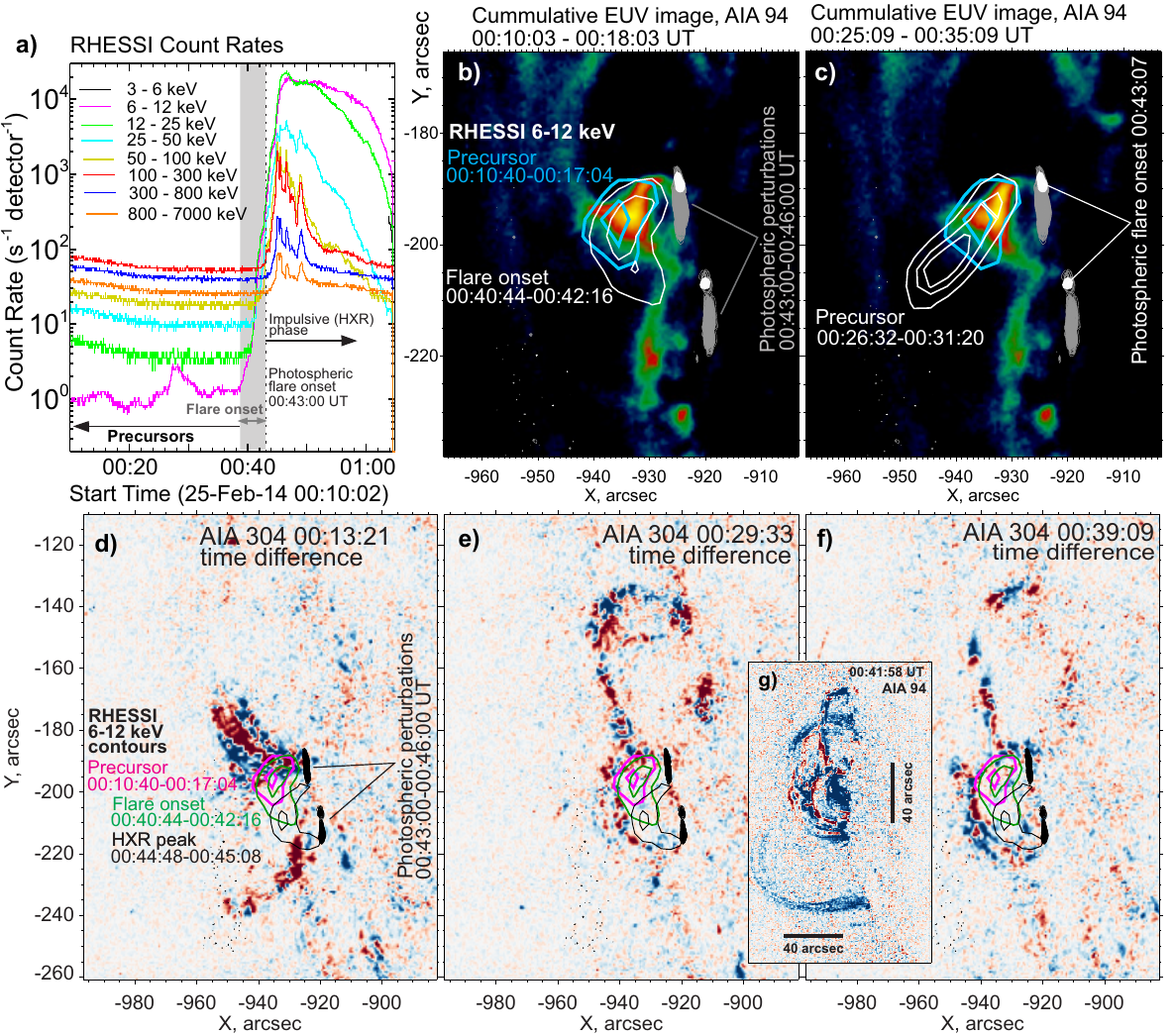}
	}
	\caption{The brief overview of the precursor phase of the selected flare. RHESSI X-ray count rates for 8 energy bands are presented in panel~a. Here we also indicate nominal flare phases (the precursor phase, flare onset, impulsive phase). In the panels b and c we compare AIA EUV cumulative image (sum of the many images during the certain time period, shown in the top of the panel) with the RHESSI X-ray 6-12 keV contour maps reconstructed for two time intervals during the precursor phase (cyan contours in both panels and white contour in c) and the flare onset (white contours in b). Grey regions marked as photospheric perturbations in panel~a and white ones named as the photospheric flare onset are places of the photospheric flare energy release found from HMI filtergrams. Red-blue running time difference AIA 304~\AA{} images in d-f show preflare energy release during the precursor phase in comparison with the X-ray sources (precursor, flare onset and the HXR peak) whose integration time intervals are plotted within the panel~d.
	}
	\label{Overview1}
\end{figure*}

In this section we will describe the general features of temporal dynamics and spatial morphology of the selected solar flare and its precursors. In Fig.\,\ref{Overview1}a we present RHESSI X-ray count rate time profiles with the three nominal (selected ``by hands'') phases: pre-flare (or, precursor) phase, the flare onset (marked by grey stripe) and the impulsive phase. 

In this research we will mostly focus on the pre-flare phase where the most energetic emission was below 12~keV and, thus, it was mainly of thermal origin. One can find a few weak gradual SXR precursors with the most intensive peak being around 00:28~UT. The flare onset is marked by the wide grey stripe beginning from the start of the SXR 6-12~keV flux rise and approximately ending at the start of the HXR 100-300~kev flux rise. It means that the flare onset time period covers thermal activation of the flare region, electrons acceleration onset (25-50~keV) and appearance of high energetic electrons (100-300~keV) far from the thermal SXR tail (usually ending approximately below 30~keV). After the onset phase we observe the impulsive phase characterizing by several HXR peaks and the highest available energy band of 800-7000~keV \citep[more details about the impulsive phase see in the work of ][]{Sharykin2023}.


The pre-flare 6-12~keV count rate was slightly above SXR background which was estimated during the night time preceding the start of the RHESSI observations of the pre-flare phase. Despite this fact, count statistics is enough to synthesize SXR images using time periods of several minutes. SXR 6-12~keV images of precursors was made for two time intervals 00:10:40-00:17:40 (cyan contours in panels b and c) and 00:26:32-00:31:20~UT (white contours in panel~c) corresponding to two SXR emission enhancements (see panel~a). In panels b and c we plotted the cumulative EUV 94~\AA{} maps (background image) obtained by summing of all 12-second AIA images within the time intervals used for the synthesis of the X-ray images. White contour in panel~b shows the position of the flare onset SXR source. Grey contours in b and c (and black ones in d-f) show the positions of the photospheric impacts forming the pair of ribbons (due to zipping motions) determined from the time sequence of HMI filtergrams. The pair of small white filled contours show positions of the photospheric flare onset triggered in the northern side of the photospheric flare ribbons.

Comparison of EUV 94~\AA{} images with the SXR 6-12 keV contour maps (Fig.\,\ref{Overview1}b-c) reveals the following observational facts. First of all, we see that the compact pre-flare SXR and hot EUV sources were co-spatial and corresponded to the loop-like morphology. The flare onset SXR emission source was also co-spatial with the pre-flare emission sources (panel b). Thus, there were clear connections between pre-flare energy release and the flare onset. Moreover the strongest flare energy release associated with the photospheric impacts was associated with the same loop-like emission source. Our working hypothesis is that we observe the approximate location of the pre-flare current sheet which lost stability and led to the flare onset.

In Fig.\,\ref{Overview1}d-f we present running time difference EUV images for AIA channel of 304~\AA{}. This channel is sensitive to low-temperatures with the peak around $10^5$~K. Here we see a complex non-stationary structure of the pre-flare emission sources. The first image revealed a cone-like ejection into the corona from the northern ribbon close to the positions of the SXR precursors. The next two panels show a nice view of two crossed EUV arcs following TCMR geometry with the pre-flare and the onset SXR sources located around the crossing.

In the following section we will investigate processes around the hot pre-flare emission sources located in the vicinity of the crossed large-scale EUV loops seen in the AIA 304~\AA{} channel. As we assume the current sheet is located somewhere in this region, we will carry the additional analysis that helps us to make a more accurate localization of the current sheet and recognize magnetic reconnection process. Generally, we are unable to see the current sheet in the solar corona directly and to measure magnetic reconnection as in the case of the Earth's magnetospheric tail when spacecraft pass through it and measure all plasma parameters in-situ. In order to find the current sheet, to define it properties, and to trace magnetic reconnection manifestations we need to make the following indirect observations of the reconnecting current sheet in the solar corona:

- To compare positions of the pre-flare emission sources with the initial HXR emission sources associated with the flare onset. We assume that the first populations of nonthermal electrons were generated in the same region where we observed pre-flare energy release;
- We need to find structures resulted from the magnetic reconnection in the pre-flare current sheet. It is not clear whether these structures are dynamic or static;
- To find changes of the photospheric magnetic field associated with the magnetic reconnection during the flare. We do not expect strong changes during the pre-flare time;
- To identify plasma outflows produced by magnetic reconnection;
- To find hot plasma with temperature approximately above 10~MK (in X-ray and EUV ranges). \\


{\bf 2. Observational results} \\

{\bf{\it 2.1. Analysis of Ultraviolet Images from AIA/SDO}} \\

{\it 2.1.1. EUV Precursors with Flare Onset Emission Sources} \\

\begin{figure*}
	\centering
	\centerline{\includegraphics[width=1.0\linewidth]{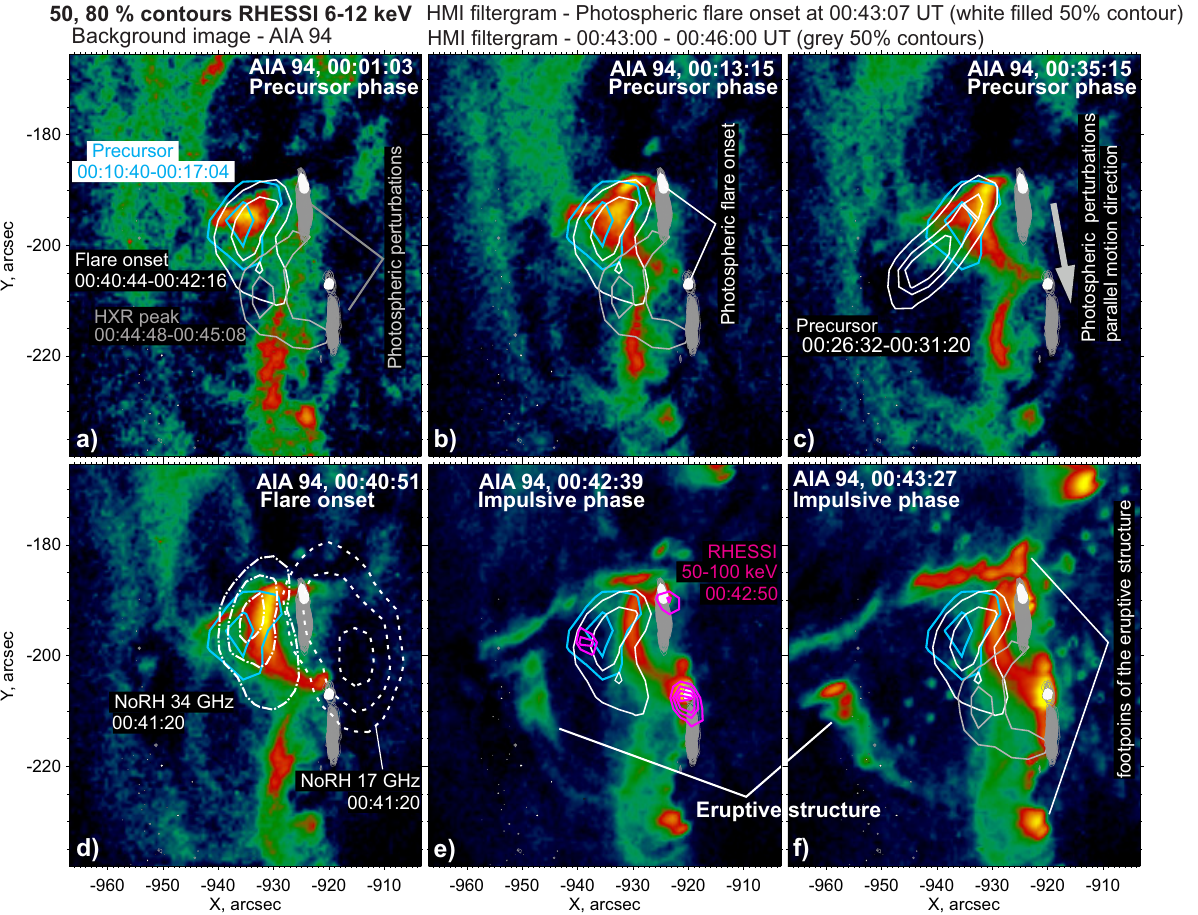}}
	\caption{Background images are the AIA 94~\AA{} maps. Cyan contours mark the X-ray 6-12 keV images reconstructed for the time interval 00:10:40-00:17:04~UT. White and grey contours in panels a-c and e-f correspond to the X-ray 6-12 keV sources for the other time ranges (flare onset, another precursor, HXR peak) shown within plots. RHESSI 50-100~keV X-ray contours are shown by magenta in panel~e. NoRH contour intensity maps for 17 and 34~GHz are plotted in panel~d. Grey regions marked as photospheric perturbations in panel~a and white ones named as the photospheric flare onset are places of the photospheric flare energy release found from the HMI filtergrams.}
	\label{Overview2}
\end{figure*}

In the previous section we discussed a few EUV images showing precursors in comparison with the SXR and photospheric emission sources. In this section we will analyze EUV maps in different AIA channels in more details revealing magnetic reconnection in the frame of the tether-cutting model. In Fig.\,\ref{Overview2} we show usual EUV 94~\AA{} images with the overlayed X-ray contour maps in different energy bands. There are three panels (a-c) showing pre-flare times, one panel~d corresponds to the flare onset and the last two ones (e-f) are made for the impulsive phase with the moving eruptive structure. 

Cyan contours in each panel show position of the pre-flare SXR 6-12~keV source generated for the time interval 00:10:40 - 00:17:04~UT. One can see that the initial EUV brightening (panel~a, 00:01:03~UT) was co-spatial with the SXR pre-flare source that is similar to the case of the cumulative EUV 94~\AA{} maps plotted in the Fig.\,\ref{Overview1}b-c. Then, we observe the EUV loop (b-c) with the loop-top which is slightly below the pre-flare SXR source. Around the flare onset (d) we see clearly the EUV loop with the footpoints in the vicinity of the onset photospheric emission sources (white filled contours). In this panel we also plot the microwave 17 and 34~GHz NoRH images for the flare onset. The high-frequency source was at the loop-top position while the low-frequency source was around the footpoints. There is no ideal co-alignment between NoRH and AIA images, but we sure that 17~GHz emission comes mainly from the footpoint region with the higher magnetic field values.

The main conclusion from Fig.\,\ref{Overview2} is that we have found clear spatial correspondence between the pre-flare energy release site and the onset flare energy release site. In other words we are sure that the flare onset started in the pre-flare magnetic configuration assocuaied with the different pre-flare emission sources. \\

{\it Integrated EUV Maps} \\

\begin{figure*}
	\centering	
	\centerline{\includegraphics[width=0.85\linewidth]{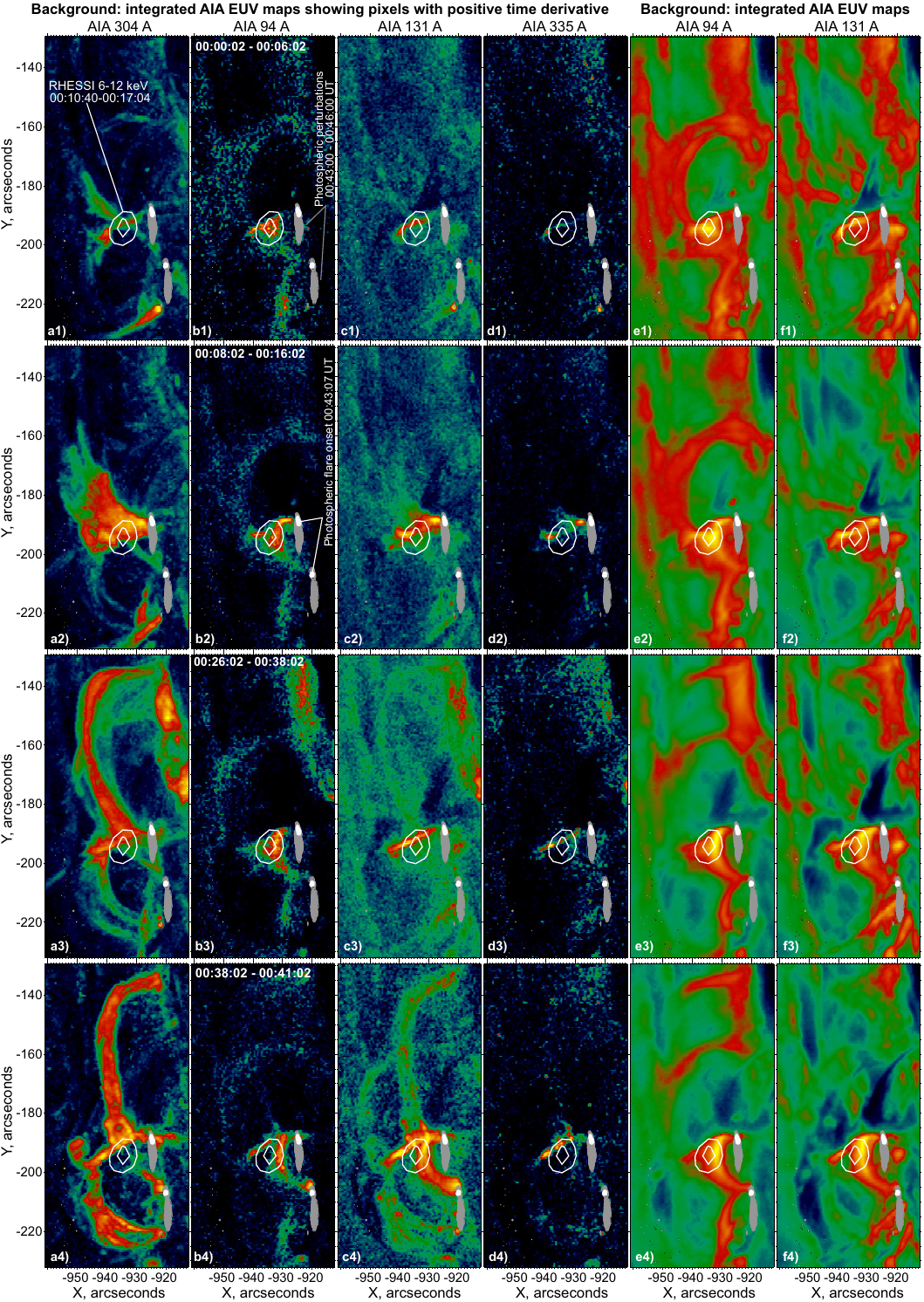} }
	\caption{This figure shows two variants of the cummulative EUV images made for different AIA channels (94, 131, 304, and 335~\AA{}) and different time intervals covering the precursor phase (a1-f3) and the flare onset (a4-f4). The first variant of the cummulative EUV map (a-d columns) is a result of summing of the maps with positive time derivatives of the EUV emission in each pixel. The other two columns (e-f) show simple integrated images (summarizing AIA 94 and 131~\AA{} images within a particular time interval). Grey regions marked as the photospheric perturbations in panel~a and white ones named as the photospheric flare onset are places of the photospheric flare energy release found from the HMI filtergrams. White contours (50 and 70 \%) correspond to the X-ray 6-12~keV map reconstructed for the time interval during the precursor phase.}
	\label{Cummulative_AIAmaps}
\end{figure*}

To analyze weak energy release in the pre-flare phase one can use time integration to make cumulative EUV maps with the reduced noise. However, we will use additional approach to achieve more sharp images containing information about the energy release rate of the precursors. All images are co-aligned with the compensation of the solar rotation. Thus, we can calculate time derivative of the lightcurve for each pixel and select only those pixels where we have positive time derivative. In this case we will consider time sequence of the maps showing spatial distribution of the positive energy release rate. In other words we will have cumulative maps of the EUV variations. Individual maps are very noisy and to achieve good maps with the reduced noise and a clear view on the spatial structure of the pre-flare energy release rate we decided to sum a few frames within particular time interval. The resulted maps are shown for four time intervals in the Fig.\,\ref{Cummulative_AIAmaps}. The first four columns correspond to the cumulative maps obtained by summing positive time derivative EUV maps (from left to right 304, 94, 131, and 335~\AA{}, respectively). The last two columns show usual cumulative maps resulting from the sum of the usual EUV images in the ``hot'' AIA channels (94 and 131~\AA{}).

Time intervals were selected to cover the pre-flare phase (three upper raws) and the flare and eruption onset (bottom raw) when the eruptive structure did not yet have enough time to fly away on the scale of a few pixels (as we need to make the cumulative maps without large-scale motions). The first time interval reveals the appearance of the localized hot coronal EUV emission source (b1-f1) co-spatial with the SXR pre-flare source. In this region we found an X-shaped EUV structure (304~\AA{} channel) confirming the tether-cutting geometry where two magnetic loops are assumed to be intersected with each other. In the second raw of the panels we found the EUV loop-like structure (b2 and e2) with the footpoints around the onset photospheric emission sources and the loop-top around the SXR pre-flare source. In other hot channels (c2, d2, and f2) we found localized EUV emission knots (size around 3$^{\prime\prime}$) lying at the base of the EUV structure seen in the cold 304~\AA{} channel (a2). In the following sections, we will discuss plasma dynamics in this region in more details and will show that there were plasma flows from the hot region. The next time interval 00:26:02 -- 00:38:02~UT (third raw of the panels) covers nominal last minutes of the pre-flare phase. The subsequent cumulative maps (a4-f4) show the pre-flare activation and transition to the flare onset (we will show it in the subsequent sections). 

The most interesting feature is the pre-flare thin line-shaped region seen in the cumulative maps of the EUV variability in the hot channels (c3 and d3). This source is also co-spatial with the SXR source and located in the cusp-like region above the hot loop seen in b3, e3 and f3. Two EUV arcs seen in cold channel of 304~\AA{} intersects with each other in the region of the SXR source and thin hot region where we observed the largest EUV variability in the hot channels. This configuration also confirms the tether-cutting geometry and now we also can discuss real geometry of the current sheet possibly associated with found thin line-shaped emission region. Around the flare onset (bottom raw of the panels) we observe bright emission from the hot loop and the region where we found the pre-flare thin region of the intensive energy release. Dark regions best seen in f4 above the hot EUV sources correspond to the eruptive filament (we will discuss it in the following sections).

To sum up all descriptions of these cumulative maps we can state that we identified the approximate morphology of the EUV emission sources. We assume that the pre-flare current sheet was located in the region of the thin EUV emission source located in the intersection of the large-scale coronal magnetic loops between the dark eruptive filament and the hot loop (where X-ray emission is also observed). From our point of view, the eruptive filament and compact loop is a result of magnetic reconnection inside the pre-flare current sheet. We will verify this idea in the following sections where a detailed analysis will be done. Our particular interest will be connected with the detailed analysis of the identified EUV structures and investigation of their connections in the frame of the tether-cutting scenario. \\

{\it 2.1.2. EUV Maps around hot emission sources. Flare and eruption onset} \\

\begin{figure*}
	\centering	
	\centerline{ \includegraphics[width=1.0\linewidth]{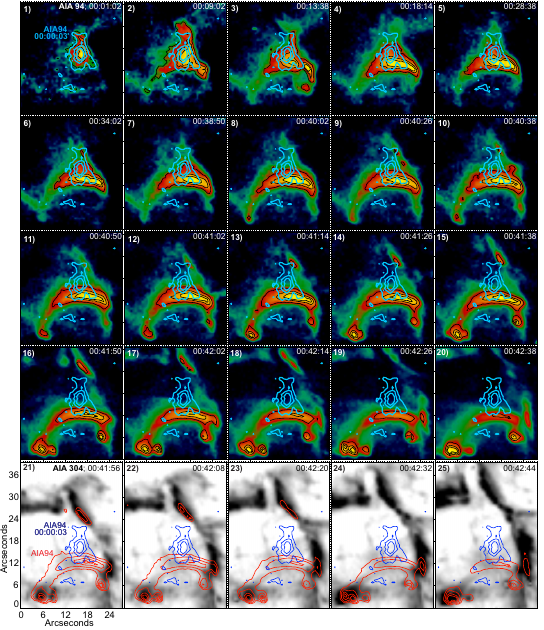} }
	\caption{This figure presents the time sequence of the AIA~94~\AA{} (1-20) and 304~\AA{} (21-25) images (rotated 90 degrees) around the loop-like structure with the pre-flare SXR source (e.g., see \ref{Cummulative_AIAmaps}-b2). The cyan contours in each panel mark the EUV 94~\AA{} emission source seen in the first time frame (00:00:03~UT) of the used data set. We show the position of this EUV source as a reference point to make an analysis of the EUV emission sources dynamics more simple. The bottom raw of the panels with the EUV 304~\AA{} maps presents the eruption dynamics comparing with the EUV hot loop seen in the 94~\AA{} channel (shown by the red contours). }
	\label{AIA_zoom}
\end{figure*}

In this section, we will discuss usual EUV images of the pre-flare hot emission sources including the thin line-shaped region and hot loop co-spatial with the flare onset emission sources. We selected a narrow field-of-view (FOV) around the hot pre-flare regions and for better visualization we rotated frames 90 degrees in the way to orient the hot loop in the vertical direction. All the analyzed images are shown in Fig.\,\ref{AIA_zoom} where we consider the AIA 94 (background in panels 1-20 and contours in 21-25) and 304~\AA{} (background in panels 21-25) channels. Data from the 304~\AA{} channel are used in this figure to show the onset and initial dynamics of the eruption. To trace vertical (radial) relative dynamics of the observed features in the EUV images we show contours (cyan in 1-20 and blue in 21-25) of the EUV 94~\AA{} emission source considered as a reference and corresponding to the initial frame (at 00:01:02~UT) of our data set.

The considered in Fig.\,\ref{AIA_zoom} time sequence of the images covers the pre-flare time interval, flare and eruption onset and the beginning of the impulsive phase when the intensive HXR emission up to 300~keV was observed. In the panels 1-5 we observe gradual appearance of the loop-like structure below the diffusive and weaker coronal emission source. Indeed in the first two frames 1-2 the pre-flare sources were diffusive without a typical loop structure. EUV maps 3-5 are likely to be a combination of the loop and the coronal cusp-like diffusive emission source (above the loop) previously seen in the first two panels. In the case of the coronal diffusive source, we possibly observe only local coronal heating around the pre-flare current sheet and energy transport along magnetic field was not so effective to initiate heating in the lower parts of the magnetic loops involved in the energy release process. The case of the loop emission source can be explained by magnetic reconnection, which most likely became more intensive and plasma in the magnetic loop below the current sheet became visible. 

The second raw (panels 6-10) shows the transition from the pre-flare phase to the nominal flare onset. Here we should comment how we understand the flare onset. First of all, we make a difference between the flare onset as a rise of different emissions intensity and eruption onset as a start of motions of magnetic structures, then transformed into a CME. No doubt, these two processes are interconnected and we will discuss it in the text below. However, the first thing that we need to do before further discussions of the physics is to find these two onset times. To catch the flare onset one can keep in mind the following facts. We know that the pre-flare emission sources and the initial flare sources were located in the same magnetic structures. Moreover, we understood the approximate morphology of the energy release site with an assumption of the localization of the pre-flare current sheet. Around the flare onset time we assume transition from the rather slow magnetic reconnection in this pre-flare current sheet to the relatively fast reconnection regime with emission intensification and sharp plasma displacements (let's consider it as the onset criteria). Thus, we have to try to find such manifestations of the activation of the pre-flare current sheet. Then, in the next sections we will make more detailed analysis of the emissions and plasma flows to verify the flare onset time.

\begin{figure*}
	\centering	
	\centerline{ \includegraphics[width=1.0\linewidth]{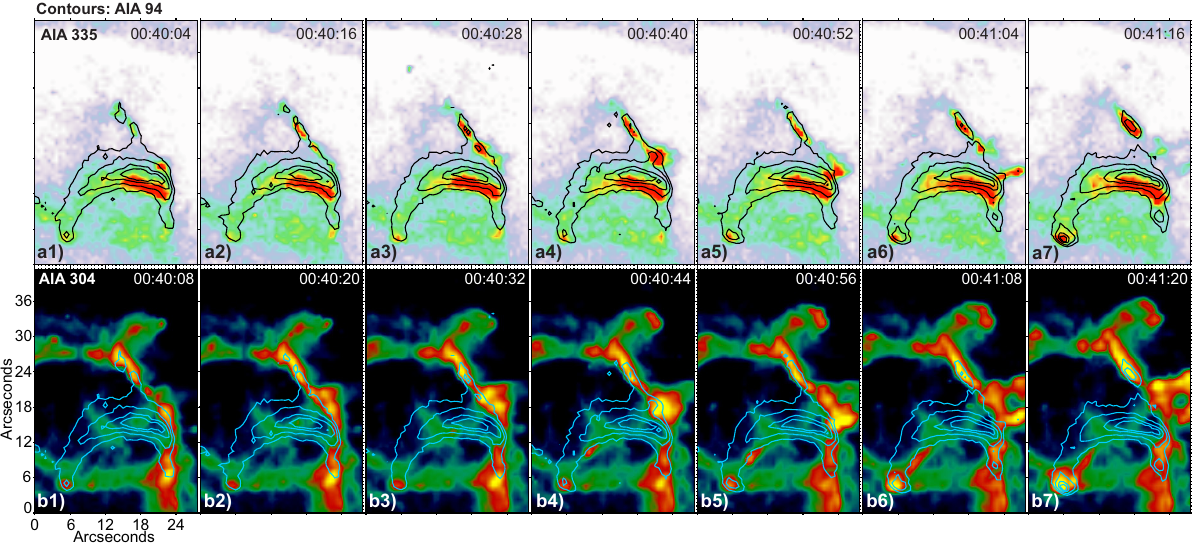} }
	\caption{In this time sequence of the EUV images we show the flare onset (triggering magnetic reconnection). Contours show the EUV source in the 94~\AA{} channel. The top and bottom raws correspond to the 335 and 304~\AA{} AIA channels, respectively. }
	\label{AIA_MRregion}
\end{figure*}

In the time sequence of the EUV images we found a few manifestations of the flare onset in the approximate time interval 00:40:02 - 00:40:26~UT according to the criteria discussed above. Firstly, there was enhancement of the EUV emission intensity in the loop footpoints (panels 8-9) and in the region of the assumed current sheet. Secondly, we found sharp plasma dynamics possibly associated with the magnetic reconnection. Let's consider these observations in more details in Fig.\,\ref{AIA_MRregion}.

Destabilization of the current sheet around the flare onset time is best seen in the 94 and 335~\AA{} channels (Fig.\,\ref{AIA_MRregion}a1-a7). One can note a relatively weak brightness enhancement and subsequent formation of two small bright moving blobs in the region of the assumed current sheet. These two hot (as we study data from the hot channels) blobs moved from each other with the characteristic velocity of 150-200~km/s (estimated from the seen displacements). Additionally, we observed in the cold 304~\AA{} channel a downward-moving blob ($\sim 100$~km/s) from the current sheet. We assume that these motions may be connected with the patchy outflows from the reconnection region. Frankly speaking, such sharp motions of a small plasma amount and weak emission disturbances cannot be a strict reason to state that we found the start of the flare. However we are sure that these phenomena were the last before all flare emissions started to grow in the regions where the pre-flare energy release took place. Such fact gives us possibility to say that the flare onset was triggered around 00:40:20~UT. More detailed analysis of the pre-flare energy release temporal dynamics in the next sections will support our findings.

As for the eruption onset, here we just have to find initial weak displacements of the eruptive filament seen in the time sequence of the EUV images. We demonstrate initial eruption dynamics in the panels Fig.\,\ref{AIA_zoom}, 11-15. We found the eruption onset in the time period 00:41:14 - 00:41:38~UT (Fig.\,\ref{AIA_zoom}, 13-15) as we see slight displacement of the coronal emission source (relative to the cyan contours). The eruption onset was also co-temporal with sharp increase of the EUV emission from the loop footpoints (possibly appeared due to nonthermal electrons) and slight downward motion of the loop apex (also relative to the cyan contours) usually considered as the loop shrinkage \cite[e.g.,][]{Somov1997,LiGan2005}. The observed eruption and shrinkage in the opposite directions from the assumed current sheet gives additional proof that our understanding of the pre-flare and flare onset current sheet location is correct. 

In the panels 16-25 we found more pronounced motion of the eruptive structure and evident loop shrinkage. For the cold channel 304~\AA{} (panels 21-25) we see clear formation of the future CME core in the form of the $\Omega$-shaped kinky structure. Important thing that this core started to form in the close vicinity of the initial current sheet where the flare onset was triggered. 

The flare onset was due to some small-scale magnetic reconnection taking place in the pre-flare current sheet that was formed long before the flare. Moreover, the small micro-flare-like reconnection preceded to the eruption onset. It gives us the reason to assume some possible resistive instability of the kinked flux rope. \\

{\it 2.1.3. Ultraviolet photometry of different EUV structures} \\

\begin{figure*}
	\centering
	\centerline{ \includegraphics[width=1.0\linewidth]{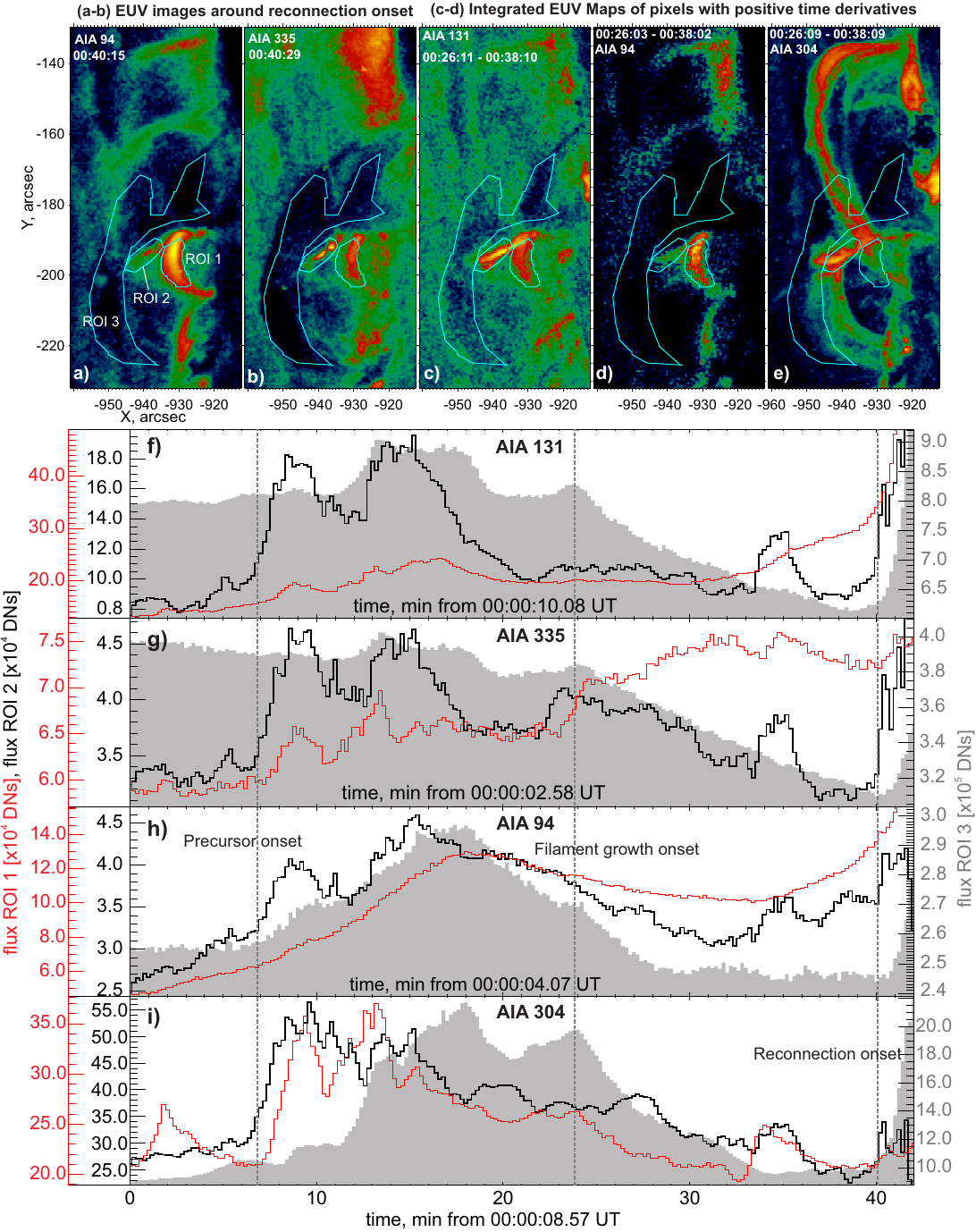} }
	\caption{The EUV time profiles for different three ROIs and AIA channels (94, 131, 304, and 335~\AA{}) are presented in this figure in the panels (f-i). The ROIs are shown by contours in the panels (a-e) where the background images are the EUV pre-flare images (a-b) and cumulative EUV maps with positive time derivatives (c-d) for the particular AIA channels and time periods during the precursor phase. The ROIs 1, 2, and 3 correspond to the loop-like EUV structure (red curves), expected magnetic reconnection site (black curves), and the location of the eruptive filament (gray shaded area), respectively. }
	\label{ROI_TPs}
\end{figure*}

To make more serious statements about the onset time of the flare and the eruption we need to study lightcurves from the pre-flare energy release region. Let's discuss temporal dynamics of the EUV emission fluxes from different AIA channels. In Fig.\,\ref{ROI_TPs} we show positions (shown by contours) of the selected regions-of-interest (ROIs) in the top panels (a-e) and the corresponding lightcurves in panels (f-i). These ROIs were selected according to the preliminary analysis of the time sequence of the EUV images. The first ROI covers the hot EUV pre-flare loop located below the assumed current sheet encircled by the contour of the ROI~2. The eruptive filament (best seen in the 335~\AA{} channel in the panel~(b) is located above the current sheet and is covered by the largest ROI~3. Along with the usual EUV images (a-b) we show cumulative maps with the positive EUV flux time derivatives (c-e). One can note that the dark filament lied above the intersecting large-scale magnetic loops (panel~e) forming the tether-cutting configuration.

Let's discuss the time profiles constructed. Despite the complex profiles, one can distinguish a few important time instants characterizing pre-flare dynamics and the transition to the flare onset. As for the hot channels (94, 131~\AA{}), the less smooth and most non-stationary spiky time profile of the EUV flux (black curves) was associated with the ROI~2, where the assumed current sheet was located. The appearance of the intensive EUV bursts was around 00:07:00~UT (the left vertical dashed line across the panels (f-i). Around this time, we detected transition from the single diffusive emission source to the loop-like emission source (see previous section). For the ROI~1 we see more gradual time profiles (red), which is consistent with the inertial chromospheric evaporation mechanism. Due to spiky energy release inside the current sheet, energy is injected into the reconnected magnetic flux tubes forming the observed EUV loop. Then the energy fluxes (e.g., heat transfer or nonthermal particles) overheats the chromosphere and induces plasma flows feeling the loop volume. This process is rather smooth due to hydrodynamics processes.

Another important feature of the time profiles is the emission decay in the ROI~3 (associated with the eruptive filament) beginning at 00:24:00~UT (the middle vertical dashed line) and best seen for the channels 131 and 335~\AA{}. It means the growth of the eruptive filament mass which is observed as darkening of the region in the particular EUV wavebands. We think that this process is also connected with the magnetic reconnection in the current sheet below the filament. The possible scenario could be as following. The new large-scale magnetic flux tubes, which likely escaped from the reconnection region, moved upward and delivered magnetized plasma into the filament's body. Due to efficient cooling, it resulted in growing of the darker regions in the particular channels. Additional plasma could also be delivered by gradual chromospheric evaporation into a larger volume of the magnetic tubes. These discussions are very qualitative and we need additional analysis of plasma motions. In the next section we will discuss the dynamics of this filament and plasma motion in more details in the context of magnetic reconnection using the time-distance plots.

The right vertical dashed line marks the flare onset around 00:40:10~UT. This time is close to those one that was estimated from the time sequence of the EUV images (see the previous section). This time instant is determined as a fast increase of the EUV emission from the current sheet region (ROI~2). One should note that this emission growth was the fastest and the most spiky especially for the channels 131 and 335~\AA{}. This time also corresponds to the start of the fast intensity growth from the ROI~3.

To sum up, analysis of the ROIs' emission fluxes allowed us to find the exact time of the flare onset and start time of the filament growth. Intensive filament growth lasted for about 15 minutes with the gradual rise of emission intensity from the EUV loop. One of the possible (from our point of view, the most likely) explanations is the gradual magnetic reconnection in the pre-flare current sheet. In the following sections, we will explore the idea of magnetic reconnection from the point of view of reconnection outflows and bulk plasma motions initiated by energy release around the reconnection region. \\

{\it 2.1.4. Eruptive Filament Dynamics: formation, growth, eruption onset} \\

One needs to study the dynamics of the filament in more details to verify the approximate eruption onset time found from the visual inspection of the time sequence of the EUV images (see above). We also need additional information about the filament growth that we found from the ROI lightcurves (in the previous section).

\begin{figure*}
	\centering
	\centerline{ \includegraphics[width=1.0\linewidth]{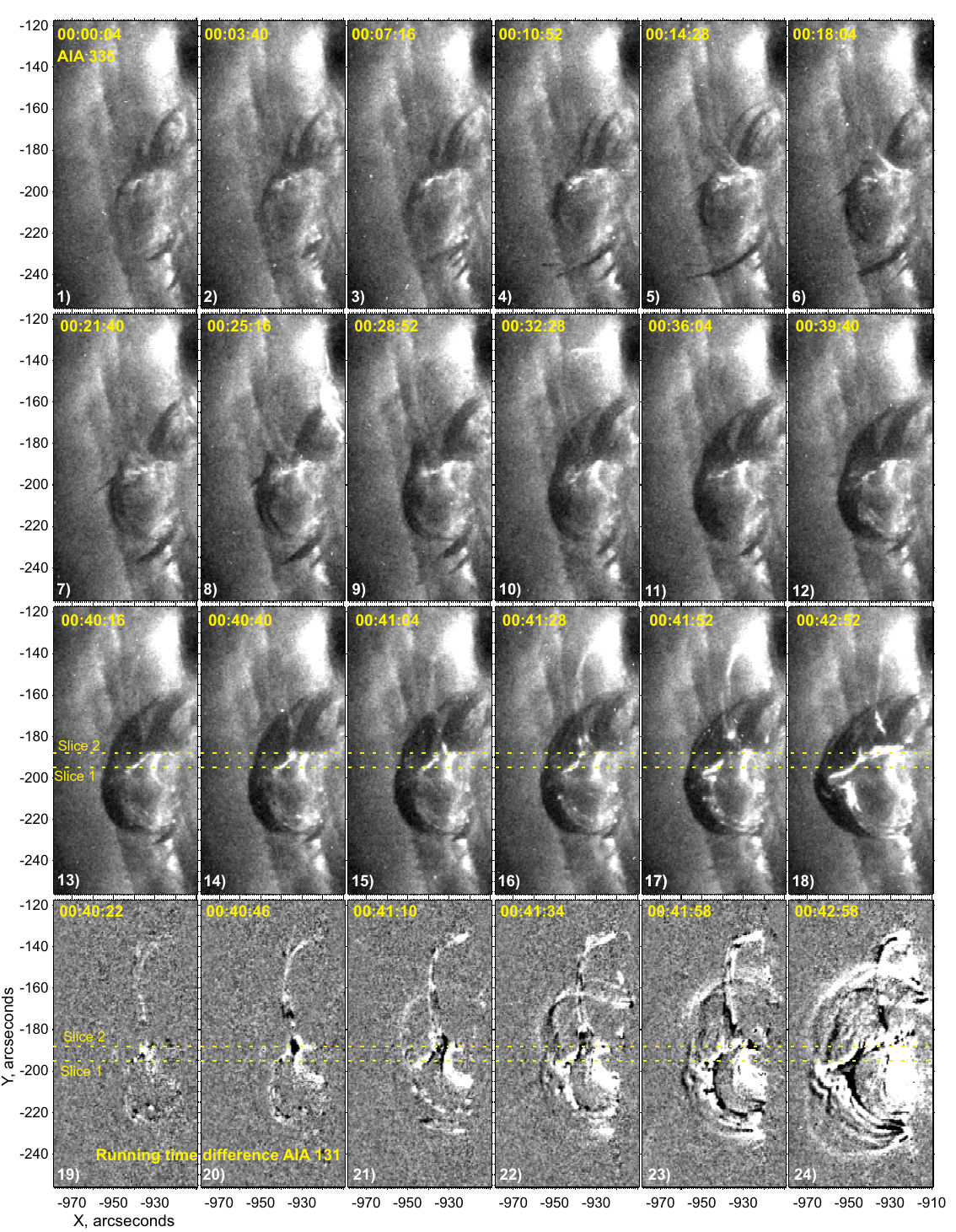}  }
	\caption{The time sequence of the AIA~335~\AA{} images is presented in the panels (1-18) to show pre-flare growth of the eruptive filament and onset of its eruption. The last raw of the panels (19-24) shows the running time difference AIA 131~\AA{} images. The horizontal dashed lines in panels (13-24) mark the observation slits used to make the time-distance plot shown in Fig.\,\ref{TD_Filament}.}
	\label{AIA304_Filament}
\end{figure*}

In Fig.\,\ref{AIA304_Filament} a time sequence of the EUV 335~\AA{} images is shown (panels~1-18). Here we have an illustration of how the filament grows. Initially, it was observed in the form of dark threads. Then it became less structured due to its growth and corresponding darkening. In the panels (7-12) one can note the bright EUV emission sources associated with the assumed current sheet and the loop. The current sheet lies exactly on the boundary of the dark filament below its apex. Precursory reconnection-like events can be recognized as dynamic jets from the current sheet observed, e.g., in the panels (5) and (6).

The panels~(13-18) show the flare region during its onset and first two minutes of the impulsive phase covering the eruption onset and appearance of the HXR emission up to 300~keV. The last panels~(19-24) present the running time difference 131~\AA{} images with times close to those ones in 335~\AA{} shown in the panels~(13-18). The frames (13) and (19) correspond to the flare onset. We see the large-scale magnetic loops intersecting below the filament body and involved in the initial flare reconnection process. The eruption onset is shown in the panels (15) and (21). Due to the motion of the eruptive magnetic structures, the outer magnetic loops forming a magnetic envelope around the filament are involved in energy release and become observable. The $\Omega$-shaped bright magnetic loop located on the bottom boundary of the filament was observable since 00:41:50~UT (panels 17 and 23). This structure will develop into the future CME core.

\begin{figure*}
	\centering	
	\centerline{ \includegraphics[width=1.0\linewidth]{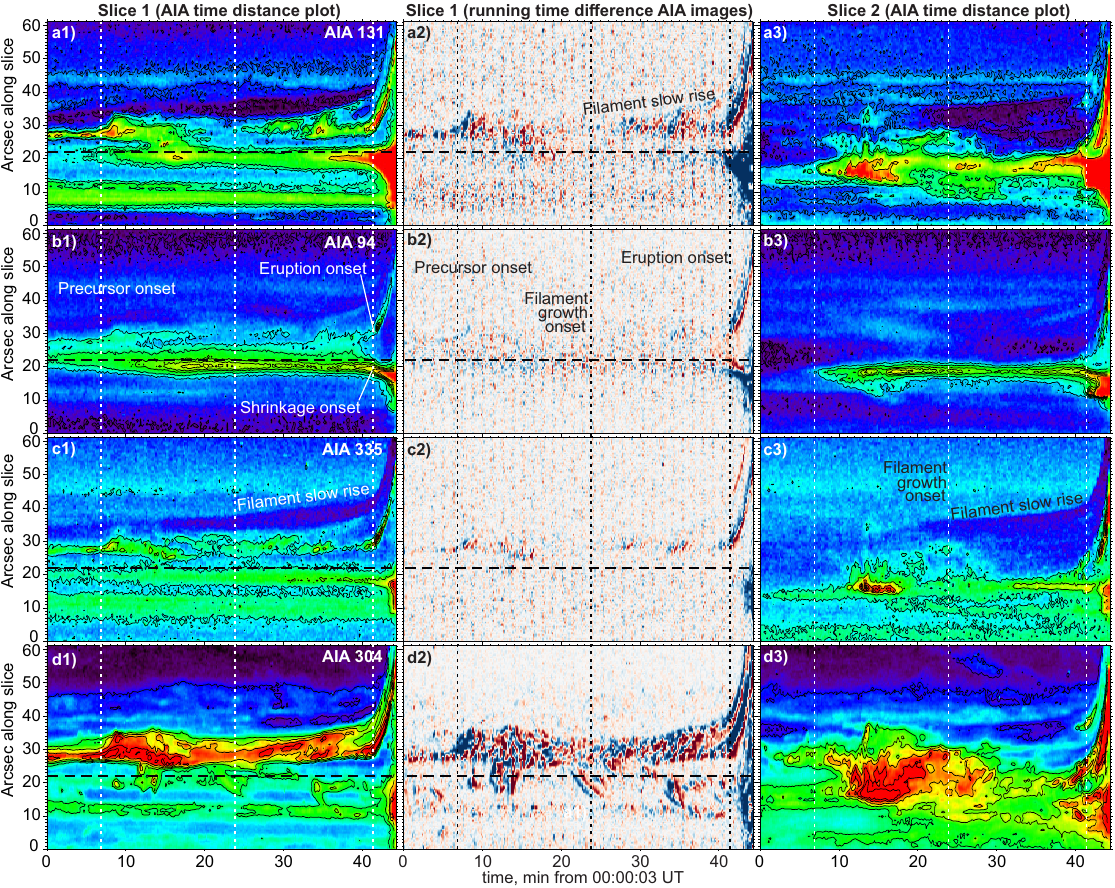} }
	\caption{Time-distance (TD) plots are presented for two observational slits shown in Fig.\ref{AIA304_Filament}-(13-24) by dashed horizontal lines crossing the dark filament, hot loop and place where the pre-flare current sheet is assumed to be located. For the slit~1 we present TD plots for usual EUV images in the AIA~131, 94, 335 and 304~\AA{} channels (a1-d1) and their running time differences (a2-d2). For the the slit~2 we show only the TD plot based on usual images in the AIA~131, 94, 335 and 304~\AA{} channels (a3-d3). Horizontal dashed line in 1-2 panel columns show approximate position of the AIA 94 \aa{} centroid (made for Fig.\,\ref{AIA_zoom}(1), cyan contours).}
	\label{TD_Filament}
\end{figure*}

To show the vertical dynamics of the eruption process in a more illustrative way we made the time-distance plots along two observational slits (two horizontal dashed lines in Fig.~\ref{AIA304_Filament}~(19-24)). These slits are located in that way to intersect the current sheet region and the filament in two places approximately corresponding to the top and bottom points of the sheet. Moreover, these slits intersect the loop-top region (slice~1) and the region where the filament's width (slice~2) was approximately the largest.

The time-distance plots are presented in Fig.\,\ref{TD_Filament} with two columns of the panels for the slice~1 and the third one showing TD plots for the slice~2. Note that the first and second columns correspond to the TD plots extracted from the time sequence of the usual (direct) images and the running time difference images, respectively. In the panels we show three time instants by the vertical lines: the precursors onset (left), filament growth onset (middle), and eruption onset (right).

During the first minutes of our data set we observed only one coronal diffusive EUV emission source in the 94~\AA{} channel (see also Fig.\,\ref{AIA_zoom}). In the TD plot we observe its slice in the (b1) panel during the first $\approx$7~min. Its centroid is marked by the horizontal dashed line, which used as a reference line to trace dynamics of the loop-top and relative plasma motions. The gradual appearance of the loop-top (best seen in the 131~\AA{} channel and shown by the dashed white line in the panel~(a1)) below the diffusive source was in the time interval 00:07:00 - 00:12:00 UT around the first timestamp (the vertical dashed line) corresponding to the activation of the precursor emission best seen in the TD plot for the slice~2 (panels a3-d3).

The filament growth onset is nicely observed in the TD plot made for the slice~2 in the form of expansion of the dark region (panels (a3) and (c3)). After the growth onset, the filament's upper point also experienced slow rise with the characteristic velocity of $\sim$3~km/s. During this expansion and rise, we observe plasma upward flows toward the filament in the form of the inclined stripes seen in the time difference TD plots (a2, c2, and d2). Their characteristic velocity lies in the range of 50-100~km/s. In the 304~\AA{} channel (d2) we also see some downward flows. These flows are very important and more details will be discussed in the next section, where we make an additional analysis. It is likely that the flows could deliver magnetized plasma to the filament, and were associated with its slow growth and preparation for eruption.

The eruption onset is best observed in the panels (b1) and (c1) in the form of a sharp break of the bright feature seen in the TD plot. Along with the eruption we also observed the shrinkage onset around the third (right) timestamp (best seen in (b1)). Oppositely directed motions of the eruptive filament and loop-top is a nice evidence for the flare onset magnetic reconnection (co-spatial with the pre-flare energy release). One can note that the flare onset in a form of small-scale magnetic reconnection was $\approx$1~min before the eruption and shrinkage onsets. It is one of the main observational findings of this work. It is likely that the eruption was triggered by the resistive mechanism. \\

{\it 2.1.5. Plasma flows around the pre-flare current sheet} \\

\begin{figure*}
	\centering
	\centerline{ \includegraphics[width=1.0\linewidth]{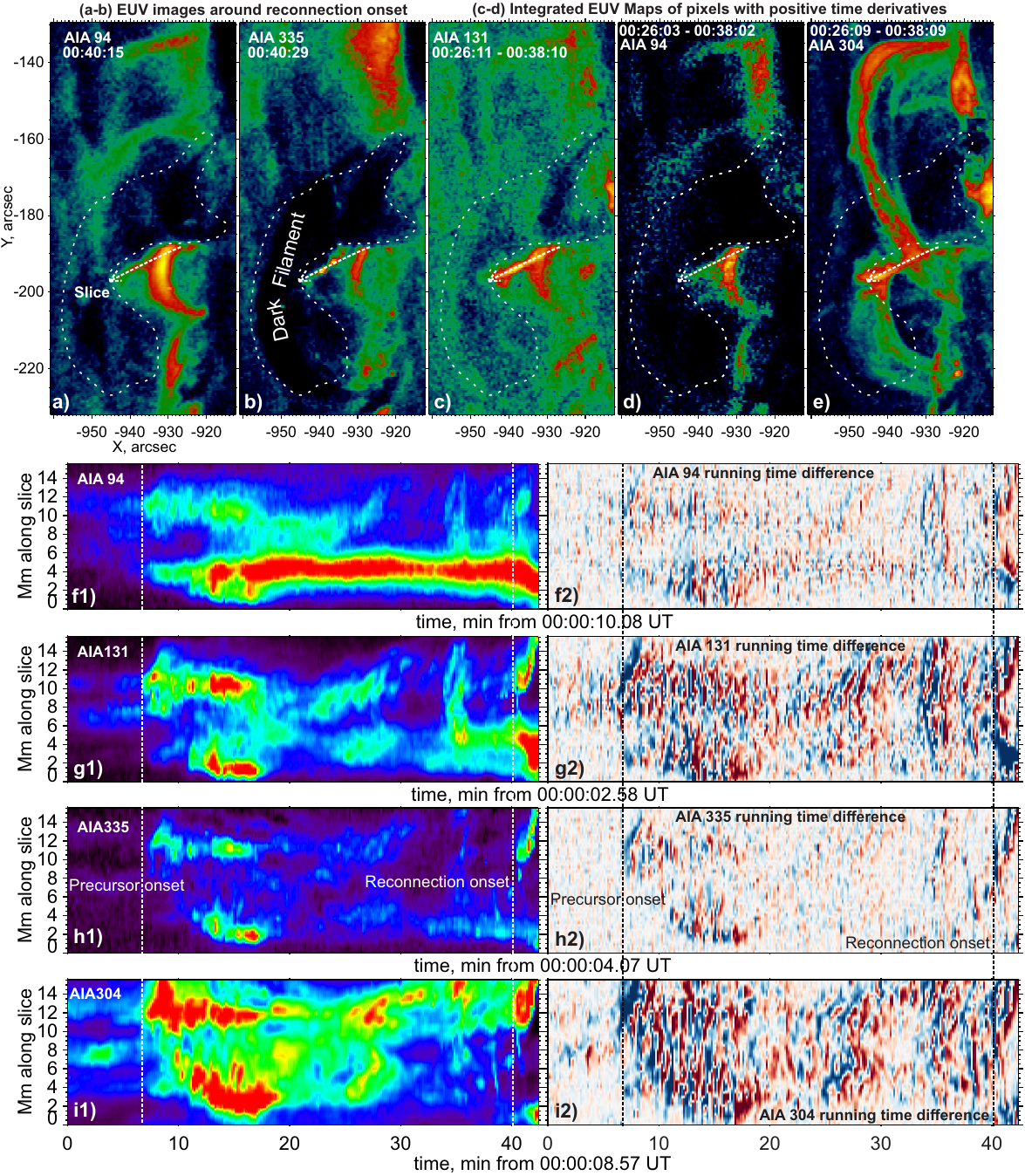} }
	\caption{This figure presents the time-distance plots along the expected current sheet. The position of the observational slit is shown by the arrow in the panels (a-e), where different EUV maps are plotted (see captions). The time-distance plots are made both for the direct EUV images (f1-i1) and for their running time differences (f2-i2) using four AIA channels (94, 131, 304, and 335~\AA{}). The dashed contour in the panels (a-e) marks an approximate observational position of the eruptive filament seen in the AIA~335~\AA{} channel.}
	\label{AlongCS}
\end{figure*}

\begin{figure*}
	\centering
	\centerline{ \includegraphics[width=1.0\linewidth]{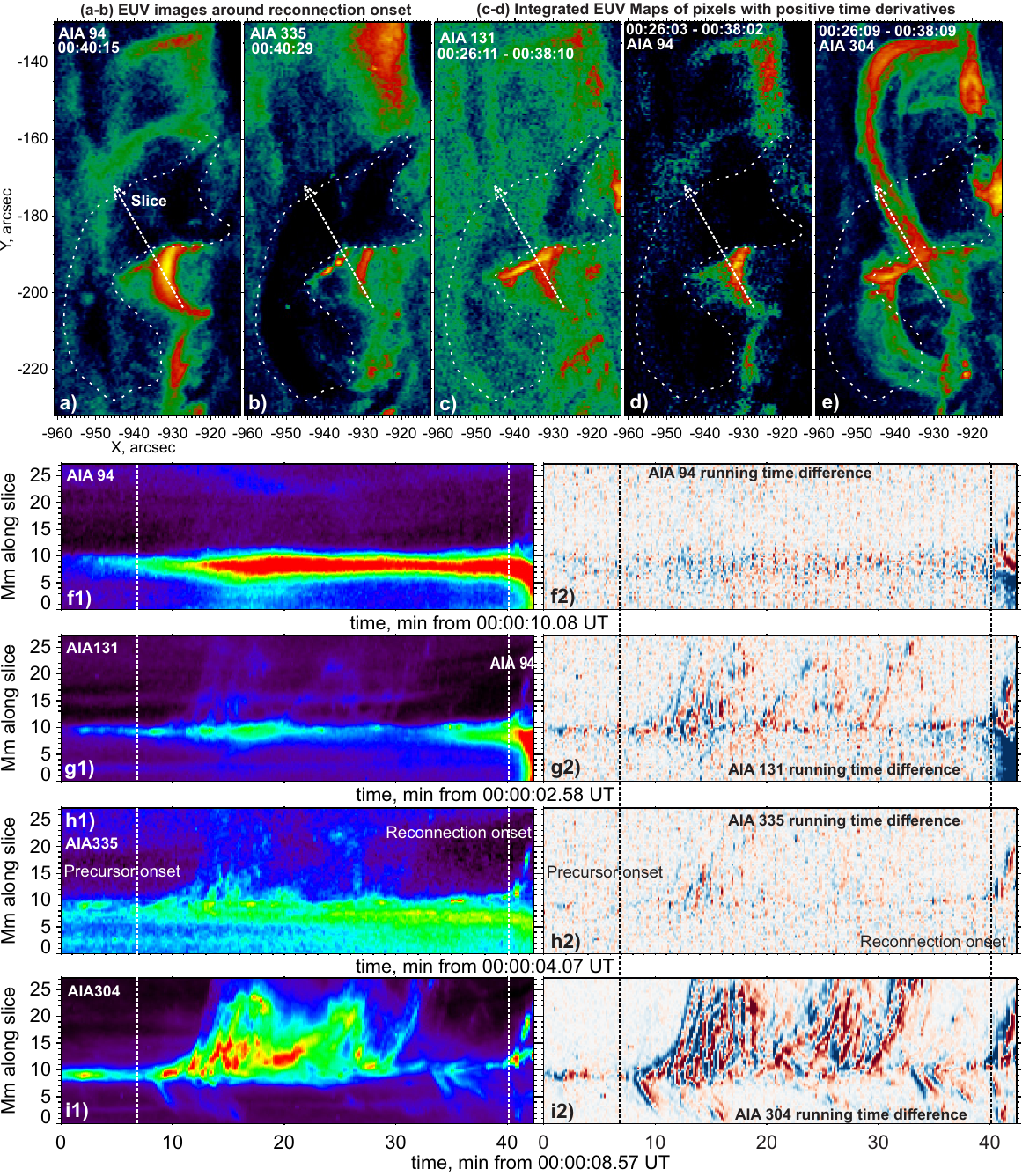} }
	\caption{This figure presents the time-distance plots across the expected current sheet. The position of the observational slit is shown by the arrow in the panels (a-e), where different EUV maps are plotted (see captions). The time-distance plots are made both for the direct EUV images (f1-i1) and for their running time differences (f2-i2) using four AIA channels (94, 131, 304, and 335~\AA{}). The dashed contour in the panels (a-e) marks an approximate observational position of the eruptive filament seen in the AIA~335~\AA{} channel. }
	\label{AcrossCS}
\end{figure*}

In the previous section, we found that there were plasma flows around the current sheet that is assumed to be in the vicinity of the thin EUV emission source. Let's analyze in more detail these flows relative to the current sheet in two directions: along and across the probable orientation of the current sheet. These directions are shown by the dashed arrows in the panels (a-e) of Fig.\,\ref{AlongCS} and Fig.\,\ref{AcrossCS}, respectively, where the usual images are in the panels (a-b) and cumulative images with positive intensity time derivatives are shown in the panels~(c-e). Note that in the Fig.\,\ref{AcrossCS}~(e) the observational slit is along the north large-scale coronal loop and potentially allows us to search plasma flows from the current sheet along the local magnetic field. In the opposite case of the slit oriented along the current sheet, we have the opportunity to see the motion of the plasma across the magnetic field lines (the slit is perpendicular to the magnetic loop).

One can see that the observational slit along the current sheet has its end close to the filament bottom. The TD plots show that there were plasma flows (seen in all panels with the running time difference images in the column (2)) along the current sheet toward the eruptive filament with the characteristic velocity $\approx$50-100~km/s. These plasma flows can be considered as the outflows from the magnetic reconnection region. The filament dynamics with its rise and expansion can be a result of the feeding by plasma from the reconnected magnetic flux tubes evacuated by outflows from the magnetic reconnection region. The most pronounced sequence of the outflows in the 131~\AA{} channel (g2) was during the filament's growth time period. It is important that we observe outflows in the hot channel as hot plasma outflows are expected from the current sheet due to magnetic reconnection energy release. After $t\approx35$~min, one can see (Fig.~\ref{AlongCS}~(g2)) episodic downward flows with the same characteristic velocity.

Flows across the current sheet, seen in the hot channels, are less pronounced and best seen in the 131~\AA{} time difference images in the form of episodic outflows directed into the coronal part of the loop. These flows were seen clearly in the cold 304~\AA{} images. Moreover, for this channel, we see clear manifestations of the flows with the same velocity in the opposite direction.

\begin{figure*}
	\centering
	\centerline{ \includegraphics[width=0.85\linewidth]{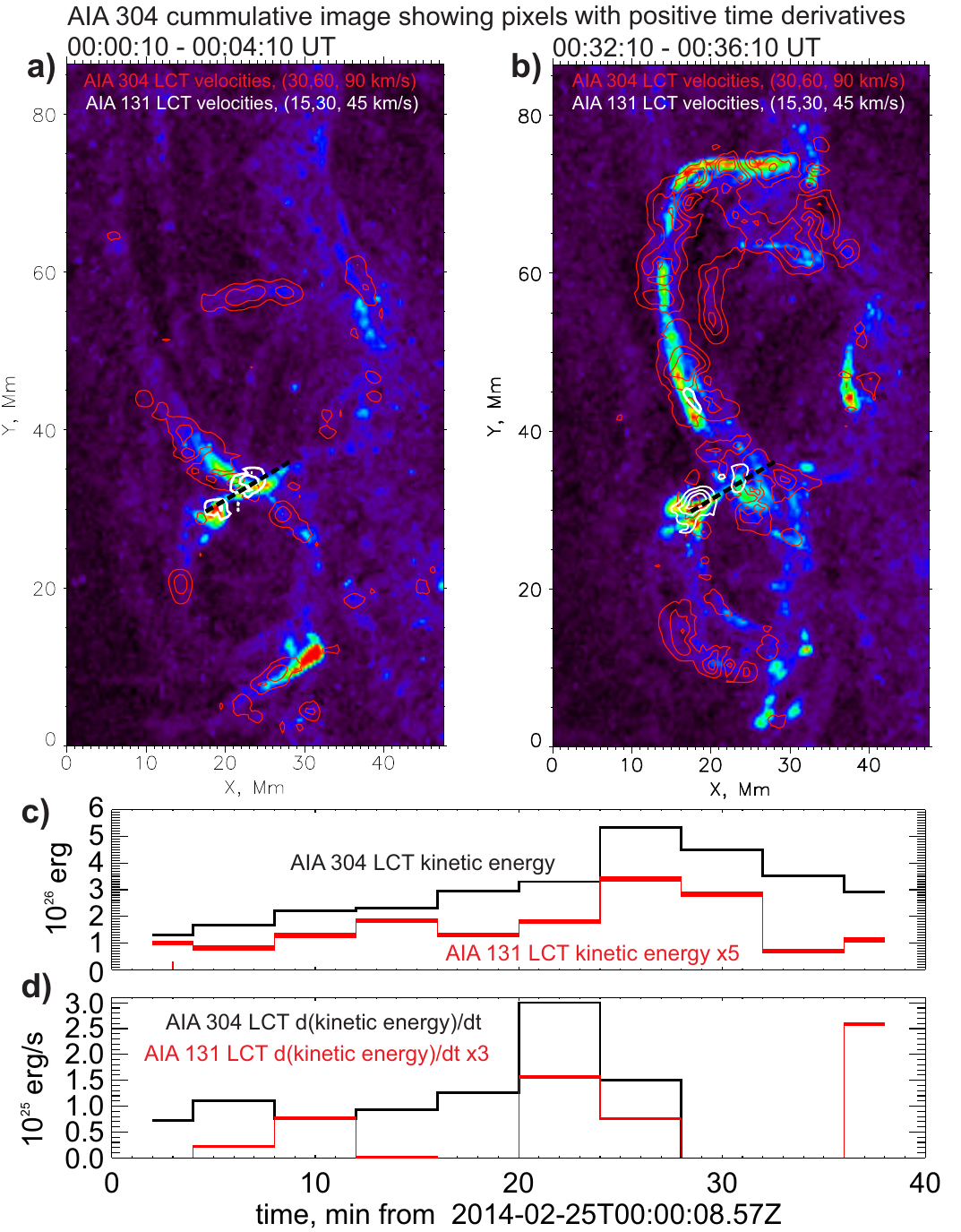} }
	\caption{These figure presents results of the LCT reconstruction of the plasma flows deduced from the time sequence of the AIA 131 (white contours) and 304~\AA{} (red contours) images for two 4-min time intervals (a, b). The contour levels are shown within the plots. The background images correspond to the AIA 304~\AA cumulative images. The black dashed line shows an approximate position of the pre-flare current sheet. The two bottom panels are devoted to kinetic energy dynamics of the plasma flows seen in these two AIA channels: (c) temporal profile of the total kinetic energy and (d) of the time derivative of the total kinetic energy above 30~km/s for the 304~\AA{} channel and above 15~km/s for the 131~\AA{} channels. For better presentation we multiplied kinetic energy values for the hot channel by 5 (the red histogram in the panel (c)) and multiplied its time derivative by 3 (the red histogram in the panel (d)).}
	\label{LCT}
\end{figure*}

For better visualization and quantitative analysis of the pre-flare plasma flows we used the local correlation tracking (LCT) technique. This approach assumes cross-correlation of many subframes in the time sequence of images to deduce velocity vector in each pixel. The resulted maps are shown in Fig.\,\ref{LCT}~(a-b) for two 4-min time intervals: (a) around the beginning of our data set and (b) approximately 4~min before the flare onset. The LCT analysis was applied to the two AIA data cubes in the hot 131~\AA{} and cold 304~\AA{} channels. It is shown by the contours with three levels, the red and white colors are for the cold and hot channels, respectively. To determine flows we used 20 images per the 4-min time interval and the subframes of 5 by 5 pixels. The most intensive flows of hot plasma were found in the localized region of the assumed pre-flare current sheet (marked by the dashed line in the panels (a) and (b)). The LCT analysis of the 304~\AA{} images reveals the large scale flow structure co-spatial with the observed crossed loops. Maximal velocity values were $\sim$100~km/s for both channels.

Using results of the LCT analysis we determined the kinetic energy (Fig.\,\ref{LCT}~(c)) of the observed flows and its time derivative (Fig.\,\ref{LCT}~(d)) in the both channels used. To determine the kinetic energy we used plasma density value about $10^{10}$~cm$^{-3}$ and the line-of-sight linear size of 3~Mm (a cross section of the loop). These values are speculative, but we think that it is enough to make a qualitative conclusion about the flows intensification, which is seen around the onset of the eruptive structure growth (00:23~UT). The peak kinetic energy value is about $5\times10^{26}$~erg and $0.6\times10^{26}$~erg for the cold and hot channels, respectively. Time derivatives have maximal values about $3\times10^{25}$ and $0.5\times10^{25}$~erg/s. However, these values should be considered only as order of magnitude estimations due to large uncertainties in linear scales and plasma density. In the following sections we compare these energy values with others (e.g., the thermal energy).

Generally, the spatial structure of the found plasma flows confirms localization of the pre-flare current sheet, reconnection in which should produce two basic types of plasma motions. We assume motions of the reconnected magnetic field lines frozen in the hot plasma and plasma flows along magnetic filed lines evacuated from the current sheet due to reconnection process (possibly due to plasma pressure gradient along magnetic field). We think that the made time-distance diagrams and LCT flow maps have clear signatures of these flows. \\

{\it 2.1.6. DEM Analysis} \\

\begin{figure*}
	\centering
	\centerline{ \includegraphics[width=1.0\linewidth]{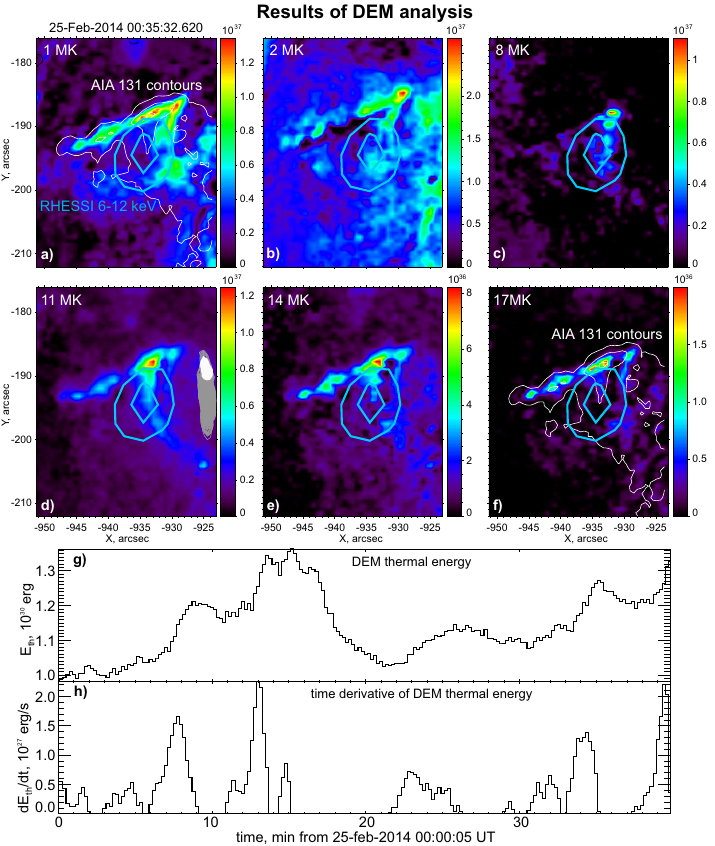} }
	\caption{This figure presents results of the differential emission measure (DEM) analysis. The panels~(a-f) show spatial distributions of EM in the pre-flare region for six temperatures from 1 to 17~MK for the single time. These DEM maps are compared with the SXR 6-12~keV source (cyan contours). The white contours in (a) and (b) reveal the EUV emission source structure in the 131~\AA{} channel. The flare photospheric disturbances are shown by the gray color in the panel (d) with the white region corresponding to the onset of the photospheric flare energy release. The bottom panels (g) and (h) show the temporal profiles of the DEM thermal energy and its time derivative, respectively.}
	\label{DEM}
\end{figure*}

Analysis of the EUV images allows us to understand general spatial structure of the energy release region and to define the qualitative picture of the temperature structure with its temporal dynamics. The AIA channels have their own sensitivities to emission of plasmas at different temperatures. To extract quantitative information we use the differential emission measure (DEM) analysis technique based on the regularization method \citep[][]{Hannah2012, Hannah2013}. Our region of interest is around the assumed current sheet. The results of the DEM analysis for this ROI is presented in Fig.\,\ref{DEM}. Six panels (a-f) demonstrate the emission measure distribution for separate temperature bins centering around 1, 2, 8, 11, 14, and 17~MK with the width $\Delta T \approx (0.5-4) \times 10^{6}$~K depending on temperature values.

The morphological structure of the obtained emission measure distributions confirms the AIA maps in different channels. We clearly see the bright linear structure and hot loop discussed in the previous sections. Moreover, we think that the resulted DEM maps visualize fine spatial structure of the energy release in better way. Indeed, we see a few (six ones in the panel (f)) emission blobs in the linear structure with characteristic spatial scale of 1-3~arcseconds. There are a few possible explanations of this fine structure. It could be due to interaction of individual magnetic loops. It is also could be connected with the tearing or thermal instability of the current sheet (see section ``Results and Discussion''). The different positions of the hot features in DEM maps comparing with the RHESSI X-ray source can be due to bad RHESSI spatial resolution and dynamic range comparing with the AIA data.  There is correspondance only around 8 and 11 MK (Fig.\,\ref{DEM}~(c-d)): similar temperatures we found from the spectral analysis (see below in the text).

The temporal profiles of thermal energies deduced from the DEM maps are shown in Fig.\,\ref{DEM}~(g-h). We used the standard approach to estimate the internal energy of the multitemperature plasma with defined DEM distributions \citep[e.g.][]{Aschwanden2015}. The maximal rise of the total thermal energy is $3.5\times 10^{29}$~ergs with the peak of heating rate $2.2\times 10^{27}$~ergs/s. Below in the text, we will compare this value with the other energy release channels. \\

{\it 2.1.7. Magnetic field in the flare region} \\

\begin{figure*}
	\centering
	\centerline{ \includegraphics[width=0.85\linewidth]{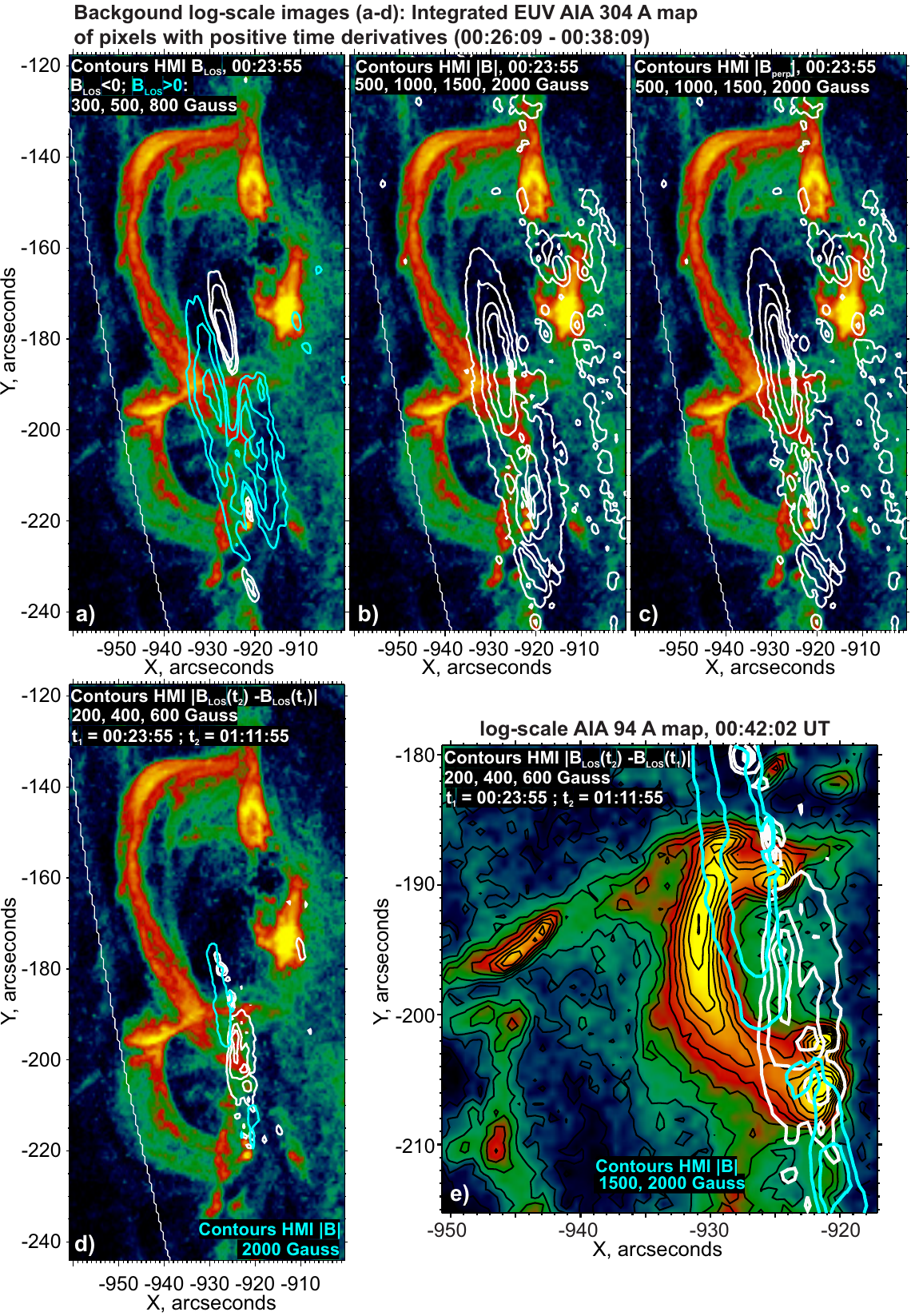} }
	\caption{The background image in (a-d) shows the AIA 304~\AA cumulative image. Contours in (a-c) correspond to distributions of the LOS magnetic field $B_{LOS}$ (a), absolute value of the magnetic field $|\bar{B}|$ (b), and plane-of-sky magnetic field component $B_{perp}$ (c). The cyan and white contours in the panel (a) are positive and negative magnetic field projections relative to the LOS field. Total flare change of the LOS magnetic field (two subtracted HMI LOS magnetograms: after the flare minus before the flare) is shown in (d) and (e) panels by the white contours; cyan contours mark absolute value of the magnetic field. The background image in the panel (e) shows the AIA 94~\AA{} EUV map, which is highlighted by contours for better contrast visualization.}
	\label{HMI_vec}
\end{figure*}

Structure of the magnetic field is highlighted by the morphology of the EUV emission sources seen in the AIA images discussed in the previous sections in details. However, it is worth to study dynamics and structure of the photospheric magnetic field. We compare the integrated EUV 304~\AA{} map with three HMI maps (Fig.\,\ref{HMI_vec}) showing distributions of the line-of-sight (LOS) magnetic component (a: $B_{LOS}$), absolute value (b: $|\bar{B}|$), and the plane-of-sky (c: $B_{perp}$) component of the magnetic field. The LOS magnetic field in the panel (a) is mostly contributed by the horizontal magnetic field component due to the projection effect (we remind that the flare was near the east limb at around S15E65). We see that the hot pre-flare loop is anchored in the inter-spot region. In this place, we observe the flare associated strongest change of the photospheric magnetic field. In the panel (d) we show position of the two sunspots (cyan color marks the 2000~G level of $|B|$) in the flaring AR and the total change of the LOS magnetic field (the pre-flare LOS map is subtracted from the post-flare one).

We conclude that the significant change of the magnetic field during the flare was in the inter-spot (probable PIL) region of two intersecting large-scale magnetic loops. Here we found that the hot loop probably associated with the magnetic reconnection in the pre-flare current sheet in the TCMR geometry. This hot loop, which is seen in the AIA 94~\AA{} channel, and the region of the photospheric LOS magnetic field change is presented in Fig.\,\ref{HMI_vec}~(e) for the time instant around the flare onset. The enhancement of the magnetic field horizontal component is produced by the loops shrinkage due to bursty magnetic reconnection during the flare impulsive phase, which is much more fast comparing with the slow regime of the pre-flare energy release process. More details about the magnetic field dynamics for this flare can be found in the work \cite{Sharykin2023} without deep discussion of the pre-flare processes. \\

{\bf {\it 2.2. X-ray emission analysis}} \\

Investigation of the hottest plasma in the pre-flare and flare region is especially important for our research as it is assumed to be directly produced by the magnetic reconnection process (e.g. heating by Ohmic dissipation, by possible shock waves). To probe such plasma one should investigate temporal and spectral properties of the X-ray emission and spatial structure of the corresponding emission sources. In this section, we will discuss X-ray observations by the RHESSI and GOES spacecraft. GOES two-channel observations are very primitive comparing with the RHESSI spatially resolved spectral data. However, it gives us standard estimations of the plasma temperature and emission measure (EM). Moreover, GOES instrument does not suffer from eclipses and covers a larger time interval in the pre-flare phase of the studied flare.

\begin{figure*}
	\centering	
	\centerline{ \includegraphics[width=1.0\linewidth]{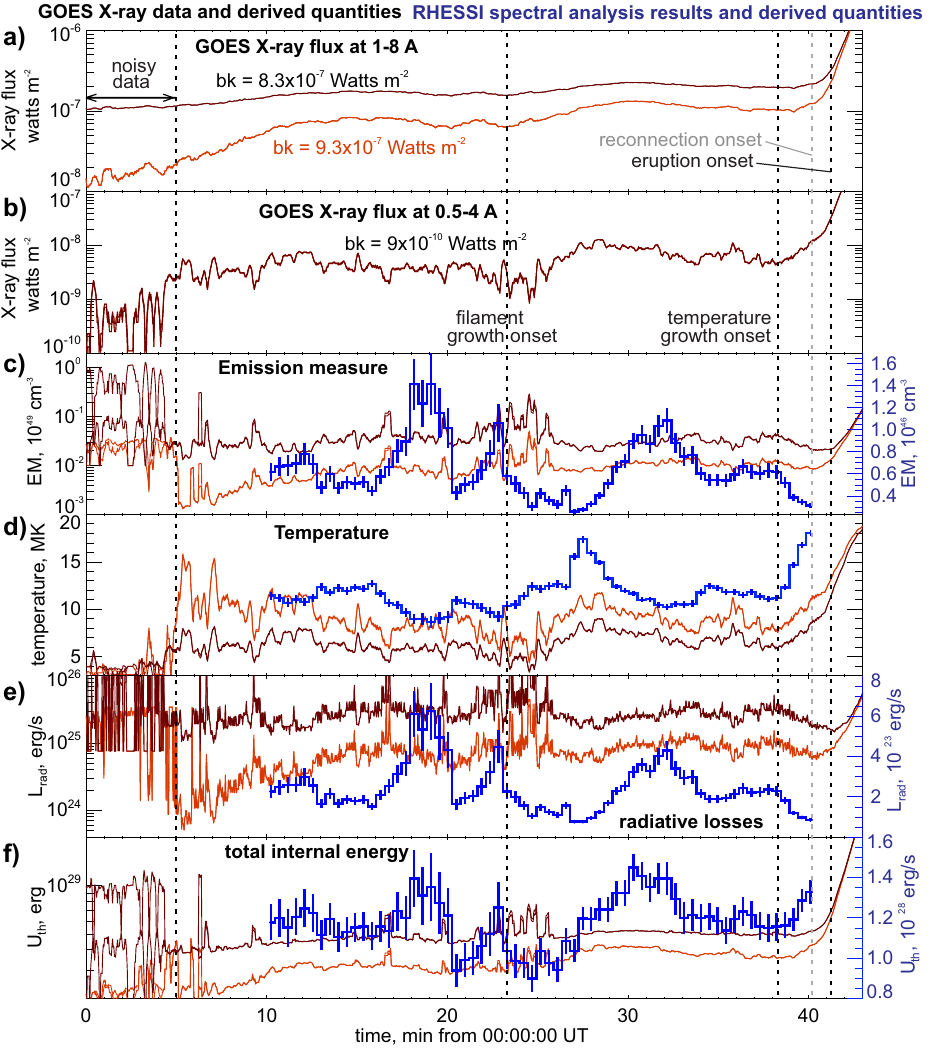} }
	\caption{This figure presents temporal profiles of the pre-flare plasma parameters and energetics estimated from the GOES and RHESSI data. The GOES 0.5-4 and 1-8~\AA{} fluxes are shown in the panels (a) and (b) for two background levels printed within these panels and marked by the brown and orange colors. The panels (c) and (d) show emission measure and temperature dynamics determined from the GOES and RHESSI data (blue histogram). The radiative energy losses and thermal plasma energy are plotted in the last two panels (e, f). 
	\label{Xray_TP}}
\end{figure*}

In the Fig.\,\ref{Xray_TP} we review the X-ray temporal profiles (a-b) with the found plasma parameters (c-d) and energies (e-f). The GOES flux curves were obtained with two background levels following technique of \cite{Ryan2012}, called as temperature and emission measure-based background subtraction method (TEBBS). This technique ensures that the derived temperature is always greater than a pre-flare background temperature and that the temperature and emission measure is increasing during the flare rise phase. As we work with the pre-flare phase with the relatively low emission fluxes comparing with the flare ones, then we need to find a quiescent background component of the flaring plasma associated with the AR emission (the whole X-ray flux of the Sun as a star will be total background). In this case, we have a background level allowing us to find a more reliable estimation of the plasma temperature and emission measure. We used GOES background fluxes at two levels of $8.3\times 10^{-7}$ and $9.3\times 10^{-7}$ Watts/m$^{2}$ for the 1-8~\AA{} channel (the total background was about $9.45\times10^{-7}$ Watts/m$^{2}$) and background flux of $9\times 10^{-10}$~Watts/m$^{2}$ for 0.5-4~\AA{} channel (the total background was about $10^{-9}$ Watts/m$^{2}$). The selected background levels give us reasonable temporal behavior of the derived temperature and emission measure. The difference between these two curves seen in the other panels characterizes the uncertainty of the derived values due to unknown background X-ray emission flux from the parent AR.

One can see that the GOES plasma temperatures differs by a factor of 1-2 and are in the range of 5-10 MK. Due to the background uncertainty the emission measure is defined with an accuracy of one order of magnitude. The temperature time profile shows the pre-heating process beginning from $\approx$00:38:00~UT. Note that we found the flare trigger burst around 00:40:10~UT. This observation is in favor of a gradual pre-heating of plasma in the current sheet before it became unstable and produced the flare onset.

\begin{figure*}
	\centering	
	\centerline{ \includegraphics[width=1.0\linewidth]{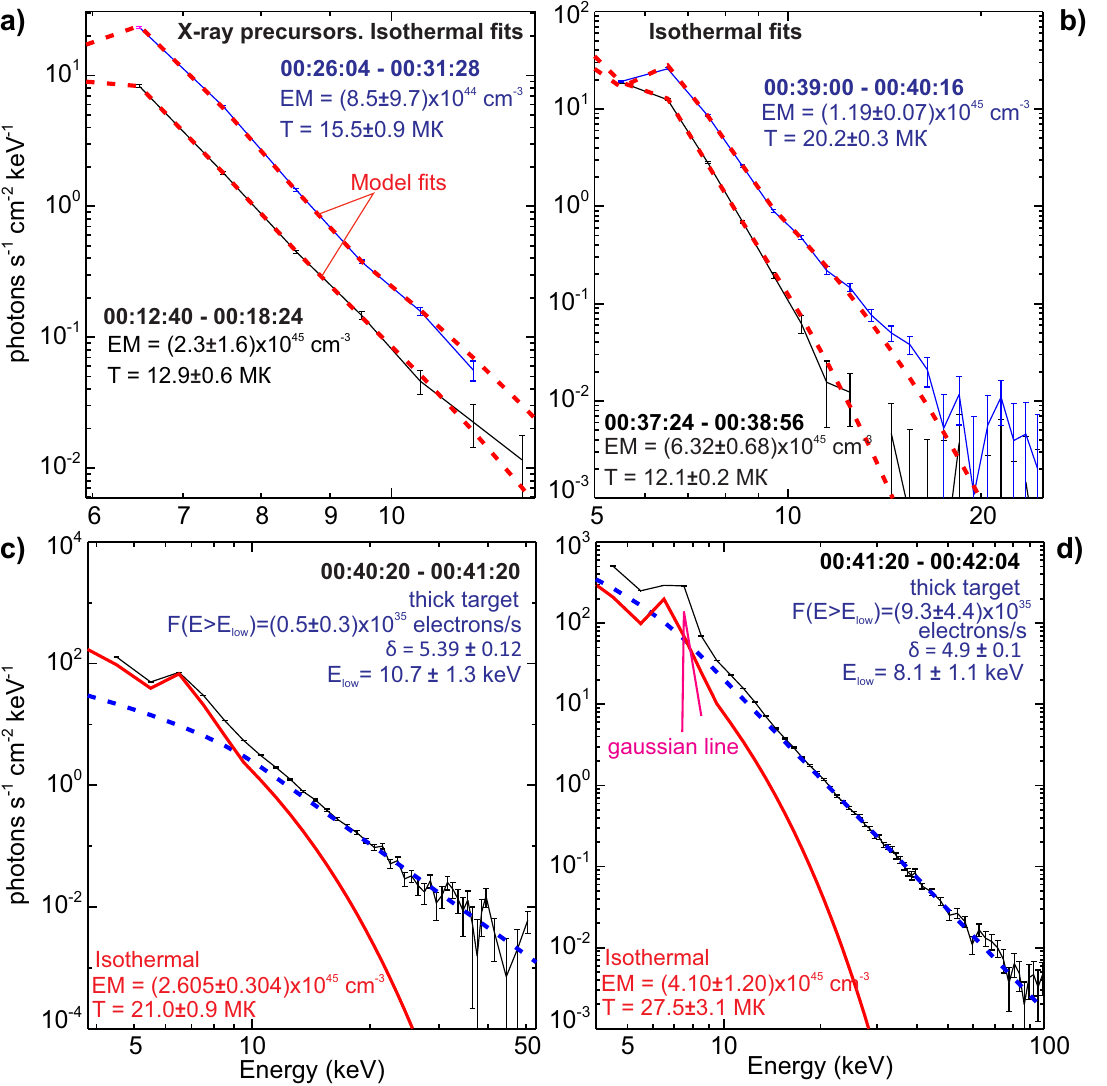} }
	\caption{Examples of the fitted X-ray RHESSI spectra. The pre-flare and pre-heating X-ray spectra are shown in the panels~(a) and (b), respectively. The X-ray spectrum shown in the panel~(c) corresponds to the time between the flare and eruption onsets. In the bottom panel, we show the X-ray spectrum after the eruption onset. The fitting models and its found parameters are written within the panels. 
	\label{RHESSI_spec}}
\end{figure*}

Results of the RHESSI X-ray analysis and energetics estimations are shown by blue color in Fig.\,\ref{Xray_TP}~(c-f). X-ray spectra were fitted by the single temperature model (using the CHIANTI atomic data-base) in the energy range of 5-20 keV using the OSPEX package. The analyzed data set ends around 00:40:10~UT (when we detected the flare onset) as in the following time intervals we need to consider the nonthermal fitting models. Since we mostly focus on the pre-flare phase, we did not make a detailed analysis of the flare impulsive phase in this paper.

The plasma temperature found from the RHESSI spectra fitting is larger then the temperature estimated from GOES observations and the temperature time profile by RHESSI varies in the range of 10-18~MK. The pre-flare temperature peak was around 00:27:00~UT and was associated with the filament growth time period. The plasma temperature time profile by GOES was more gradual, but one can also note the local temperature peak around the same time stamp. The pre-flare preheating is also observed in the plasma temperature curve by RHESSI beginning from 00:38:00~UT and rising up to 17~MK. Emission measure dynamics shows opposite behavior for the two used instruments: larger values for GOES ($EM\approx 10^{47}-10^{48}$~cm$^{-3}$) and smaller for RHESSI ($EM\approx 0.3\times 10^{46}-1.5\times 10^{46}$~cm$^{-3}$). The difference between temperature estimations by GOES and RHESSI is due to their different sensitivities to plasma temperature distributions. We mean that the wideband GOES detectors are sensitive to emission of plasmas with lower temperatures comparing with the RHESSI spectral data above $\approx$3~keV.

Flare energies found from the RHESSI and GOES X-ray data are shown in Fig.\,\ref{Xray_TP}~(e-f): the radiative losses $L_{rad}$ (e) and the total internal thermal plasma energy $U_{th}$ (f). All these energies are calculated according to the standard formulas \citep[e.g., ][]{Sharykin2015c} depending on plasma temperature, emission measure and emission source volume (only for the internal energy). We have different values of the calculated energies from the GOES and RHESSI data according to different thermodynamic parameters shown in the upper panels. The energy deduced from the GOES data is averagely larger ($L_{rad}\approx 10^{24} - 5\times 10^{25}$ and $U_{th}\approx 2\times10^{28} - 6\times 10^{28}$ ergs) comparing with the RHESSI ones ($L_{rad}\approx 10^{23} - 1.4\times 10^{23}$ and $U_{th}\sim 10^{28}$ ergs). However, it is worth noting that we have a very uncertain volume occupied by the X-ray emitting plasma. For the RHESSI data we can estimate the heating rate (e.g., between 25 and 30~min in Fig.\,\ref{Xray_TP}~(f)) as $dU_{th}/dt\approx 1.2\times 10^{25}$ ergs/s, which is smaller than the radiative losses. It is worth noting that according to the work of \cite{Hannah2011} the microflares have thermal energies in the range of $10^{27}-10^{30}$ ergs that is similar to our values found in the pre-flare time period.

In Fig.\,\ref{RHESSI_spec} we show examples of the single temperature fit of the RHESSI X-ray spectra for a few broad time intervals: precursors (a), preheating before the flare onset (b, black) and the flare-eruption onset (b, blue). We also show the X-ray spectra during the beginning of the flare impulsive phase in the panels (c-d). We consider the two-component approximation using the single temperature plasma and the thick-target model of non-thermal bremsstrahlung X-ray emission \cite{Brown1971} produced by accelerated electrons.

\begin{figure*}
	\centering	
	\centerline{ \includegraphics[width=1.0\linewidth]{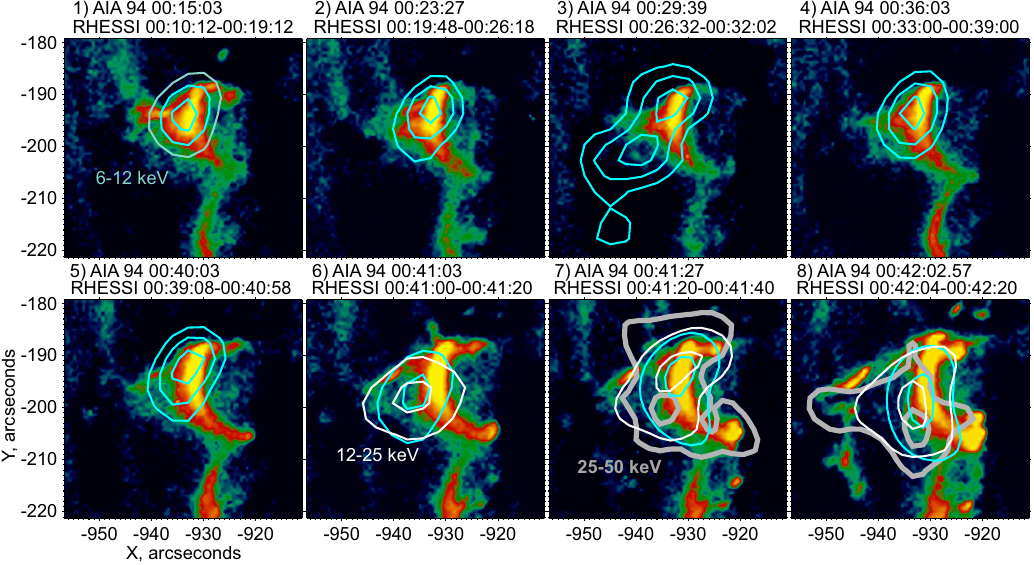} }
	\caption{The AIA 94~\AA{} images (background maps) are compared with the RHESSI contour maps in three energy bands: 6-12 (cyan contours), 12-25 (white contours), and 25-50~keV (thick grey contours). Eight frames cover the pre-flare time period and the flare onset including the reconnection and eruption onsets. \label{Xray_AIA_Images}}
\end{figure*}

In the previous sections we have shown the positions of the X-ray sources relative to the EUV emission sources seen in different AIA channels. However, it is interesting to trace dynamics of the X-ray emission sources from the pre-flare time period to the flare reconnection onset and start of the eruption process. Eight frames in Fig.\,\ref{Xray_AIA_Images} present a comparison of the hot EUV 94~\AA{} images (background colored maps) with positions of the X-ray sources (shown by contours) reconstructed in three energy bands (6-12, 12-25, and 25-50~keV). One should note that the X-ray images were reconstructed for different time intervals needed to synthesize qualitative images with the contrast emission sources. The AIA frames were selected approximately in the middle of the RHESSI time intervals used to reconstruct the X-ray images.

First four (upper raw) panels show the pre-flare time period where the X-ray 6-12~keV emission source were associated with the hot loop-top region. The panel~(5) corresponds to the preheating time period and the reconnection onset around 00:40:10~UT. This panel reveals the same position of the X-ray source as in the case of the pre-flare time period. The next frame~(6) shows the eruption onset time. We see appearance of the X-ray 6-25~keV emission source (two sources 6-12 and 12-25 keV are shown), centroids of which were located above the bright EUV loop and below the erupting structure. Beginning of the impulsive phase is shown in the two subsequent panels (7-8). We observe appearance of the coronal HXR 25-50 keV emission source above the loop and below the erupting EUV structure. In the case of the last (bottom right) panel the HXR source is elongated from the shrinking EUV hot loop toward the eruptive structure. The discussed sequence of the X-ray images of different energies clearly shows dynamics of energy release in the compact region from the relatively weak pre-flare energy release to the onset of the flare impulsive phase. It shows a causal relationship between the low intensity X-ray emission sources before the flare and the flare itself with the corresponding eruption.

\begin{figure*}
	\centering	
	\centerline{ \includegraphics[width=1.0\linewidth]{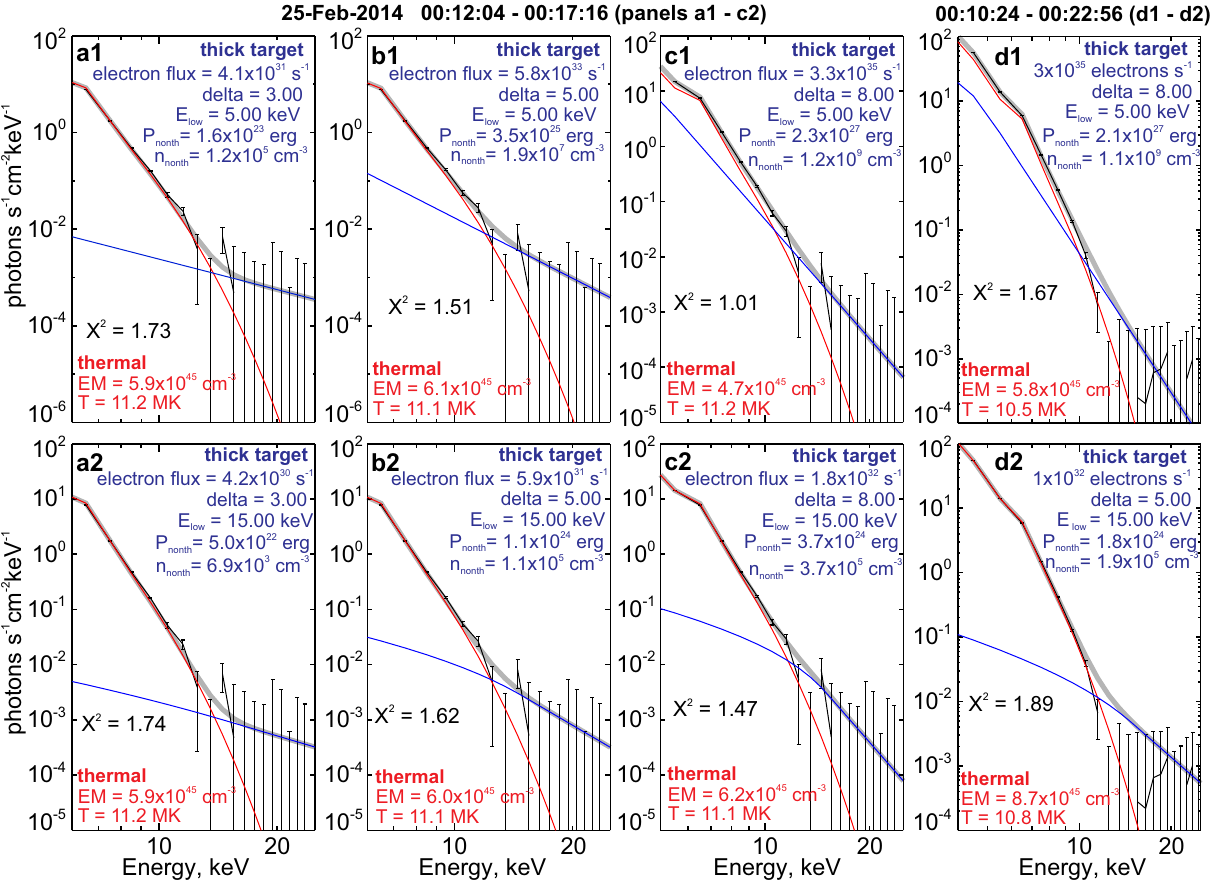} }
	\caption{X-ray spectra measured by RHESSI and results of their fitting. There are two pre-flare time intervals shown in the panels (a1-c2) ($\sim 5$~min) and (d1-d2) ($\sim 12$~min). Spectral fitting was done with the fixed low-energy cutoff (5 and 15~keV) and spectral power-law index of nonthermal electrons (3, 5, and 8). The plasma emission measure, temperature, and total flux of nothermal electrons are free fitting parameters. The thermal single-temperature and nonthermal power-law thick-target bremsstrahlung are shown by the red and blue lines, the data points (with errors) are shown by black.}
	\label{RHESSI_spec_th_nth}
\end{figure*}

Our analysis of the pre-flare X-ray emission was based on the absence of the nonthermal electrons. Indeed we do not observe evident power-law tails in the pre-flare X-ray spectra. However, we can try to estimate possible nonthermal electrons fluxes by X-ray spectra approximations using a single temperature model combining with the thick target model at a fixed low-energy cutoff and fixed power-law spectral index of the accelerated electrons. In such way we only vary emission measure, temperature of thermal plasma and total nonthermal electrons flux. To do it we selected two broad time intervals covering the pre-flare phase: 00:12:04 - 00:17:16 UT (relatively narrow) and a wider one 00:10:24 - 00:22:56 UT (check Fig.\,\ref{Xray_TP}). Results of different spectral fits are summarized in Fig.\,\ref{RHESSI_spec_th_nth}. For both time intervals we do not observe significant changes in plasma temperatures ($T=10.5-11.2$ MK) for $\delta=3-8$ and low energy cutoff ($E_{low}$) values of 5 and 15 keV. The emission measure is also indifferent to the form of the nontermal spectra and values of nonthermal electron fluxes and varies in the range of $4.7-8.7\times 10^{45}$~cm$^{-3}$. In Fig.\,\ref{RHESSI_spec_th_nth} we present values of the total electron flux ($F$) and the corresponding nonthermal kinetic power with the density of accelerated electrons estimated as $P_{nonth}=FE_{low}(\delta-2)/(\delta-1)$ and $n_{nonth}=F/S\sqrt{E_{low}/(2m_{e})}(\delta-3/2)/(\delta-1)$, where $S=0.6\times 10^{17}$~cm$^2$ (an estimated area of the loop footpoints from the EUV images), and $m_e$ - electron mass. This analysis reveals that the presence of nonthermal electrons was possible with their total energy larger than $L_{rad}$ and $dU_{th}/dt$ according to the RHESSI and GOES data at the case of a small low energy cutoffs and very soft spectra (Fig.\,\ref{RHESSI_spec_th_nth}c1-d1: $P_{nonth}\approx 2\times 10^{27}$ erg/s). We do not think that it corresponds to real situation because we have noisy data above 15~keV (due to data errors and background uncertainty), and factually we are discussing upper limits on the nonthermal spectra hidden by the thermal spectral part and noisy high-energy region. However, we cannot rule out and ignore the possible hidden nonthermal energy (comparable with the thermal one) due to the limited sensitivity of the used instruments. \\

{\bf {\it 2.3. Microwave Emission Analysis}} \\

\begin{figure*}
	\centering
	\centerline{ \includegraphics[width=1.0\linewidth]{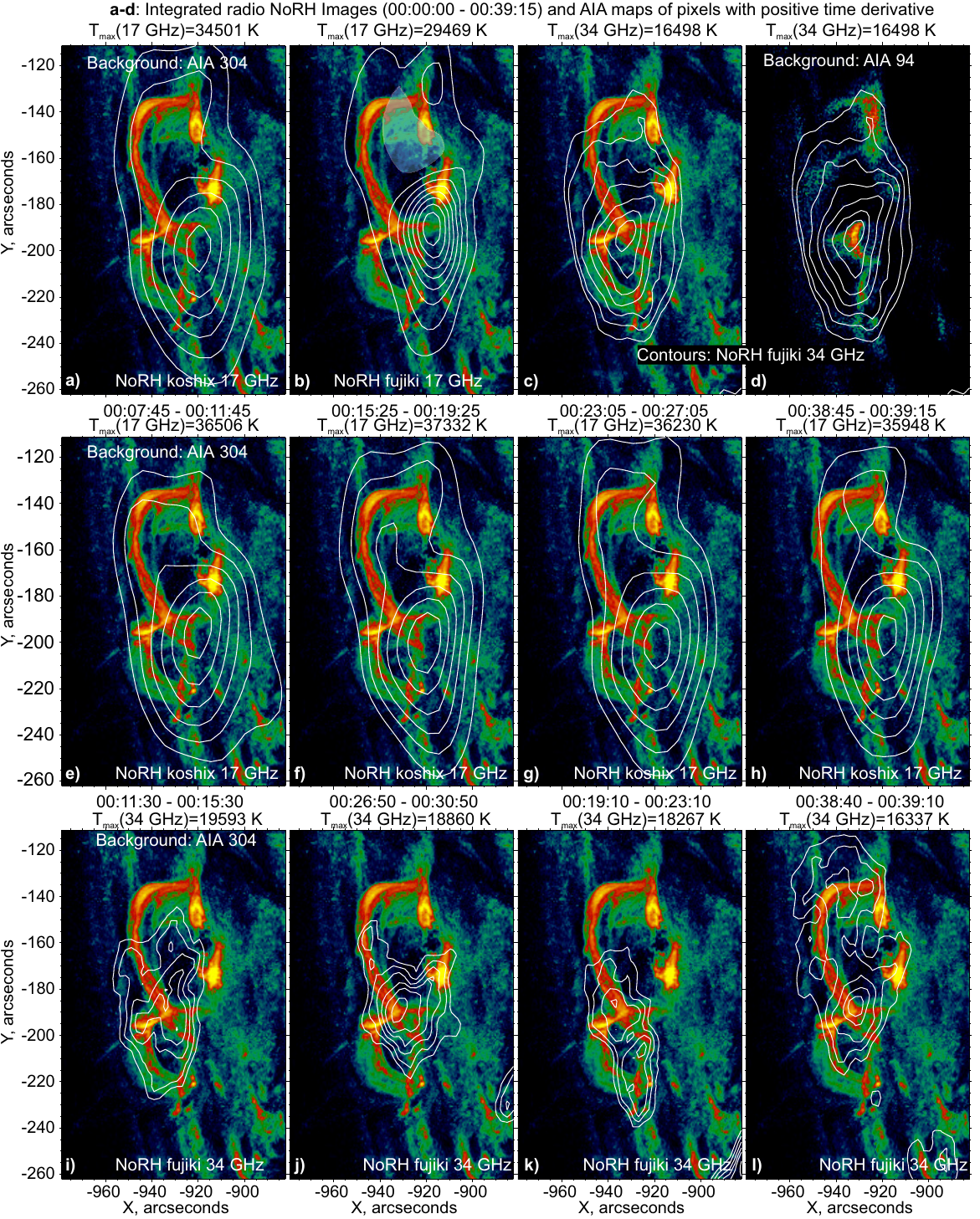} }
	\caption{The NoRH microwave image contours (white) for two frequencies 17 and 34~GHz overlayed on the AIA EUV images. Two reconstruction algorithms ``koshix'' (only at 17~GHz) and ``fujiki'' (at both frequencies) are used. The top panels (a-d) show the averaged radio images over the entire pre-flare period 00:00:00 - 00:39:15 UT studied: ``koshix'' 17~GHz (a), ``fujiki'' 17~GHz (b) and 34~GHz (c) over the cumulative AIA 304~\AA{} map, and ``fujiki'' 34~GHz over the cumulative AIA 94~\AA{} map (d). The time sequence of the averaged NoRH images overlayed on the cumulative AIA 304~\AA{} map are shown in the panels (e-l): ``koshix'' 17~GHz (e-h) and ``fujiki'' 34~GHz (i-l). \label{NoRHims}}
\end{figure*}

In this section we demonstrate results of the analysis of imaging microwave NoRH data at frequencies of 17 and 34 GHz. We used two different reconstruction algorithms: ``koshix'' (at 17 GHz, it is more sensitive to diffuse sources)  and ``fujiki'' (17 and 34 GHz, it is more suitable for discrete flare like sources). A summary of the microwave radio maps (brightness temperature distributions) comparing with the integrated EUV maps (304 and 94~\AA{}) are shown in Fig.\,\ref{NoRHims}. In the case of the selected two algorithms, in the panels (a-b) we present the total integrated 17 GHz radio maps through the broad time interval covering almost the entire pre-flare phase 00:00:00 - 00:39:15~UT. We see that the source centroids of the maximal brightness temperature $\sim 30000$~K at 17~GHz are around the pre-flare hot loop footpoints. There are likely two emission sources: the first one is connected with the compact pre-flare energy release site around the hot EUV loop and the second diffusive one is associated with the large-scale loops. This difference is clear in the case of both algorithms as we see gradients of the contour levels. These gradients are less pronounced in the case of the 34~GHZ images shown in the panels (c-d). We see that the 34~GHz centroids are around the hot EUV loop-top (94~\AA{}, where we observed the soft X-ray sources) with the maximal value $T_{br} \approx 16500$~K. The places of the microwave sources at 17 and 34~GHz do not coincide to each other, but they are evidently from the magnetically coupled regions in the pre-flare hot loop observed during all the studied time intervals.

NoRH images at 34~GHz are more noisy in comparison with the 17~GHz maps. To reduce noise we constructed the time averaged radio maps with a cadence of 4~min. ``Koshix'' 17~GHz contour maps are shown in Fig.\,\ref{NoRHims}~(e-h) and ``fujiki'' 34~GHz are in the panels (i-l). Both frequency channels reveal dynamics of the emission sources with much larger variations at high-frequency. Note that the brightest regions at 17~GHz are located in the footpoints of the pre-flare hot loop, while the highest brightness temperature at 34~GHz is around the loop-top, where the large-scale magnetic loops intersect each other.

\begin{figure*}
	\centering
	\centerline{ \includegraphics[width=1.0\linewidth]{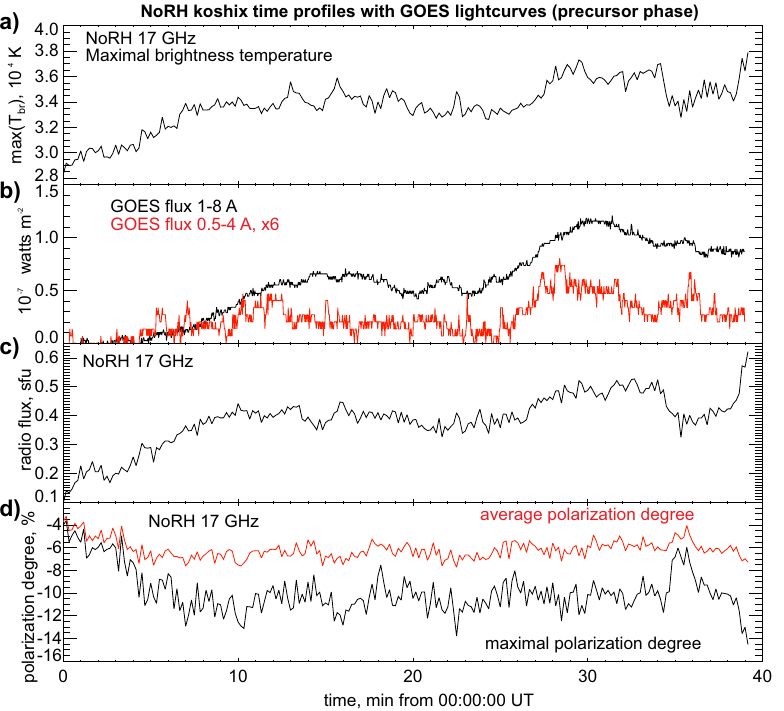} }
	\caption{Time profiles of the NoRH 17~GHz maximal brightness temperature in the pre-flare region (a), GOES X-ray flux in both channels (b),  NoRH 17~GHz total radio flux from the pre-flare region (c), and polarization degree (d) in the averaged case (red) and the maximal absolute value (black) over the pre-flare region. \label{NoRH_TPs} }
\end{figure*}

\begin{figure*}
	\centering
	\centerline{ \includegraphics[width=0.9\linewidth]{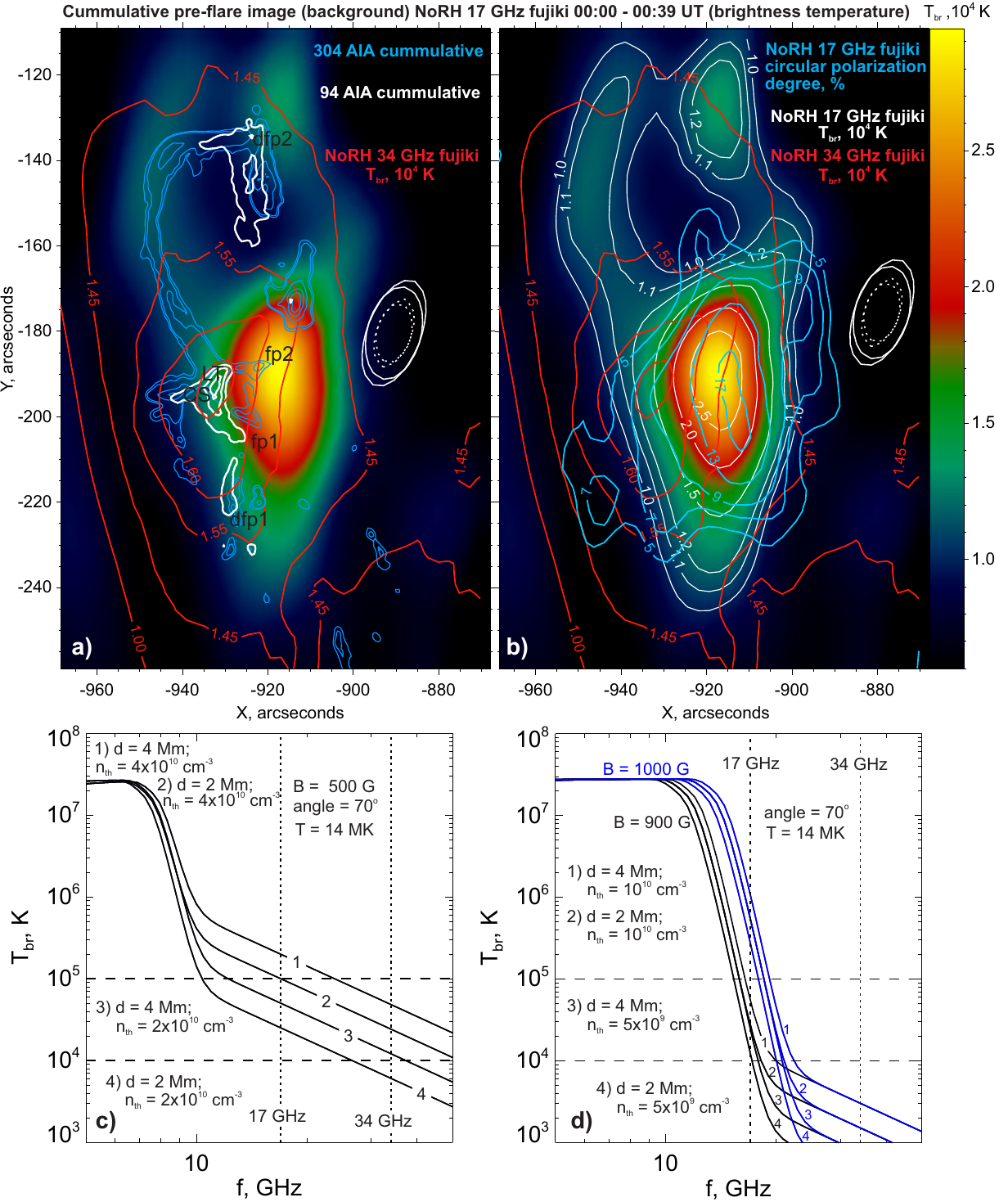} }
	\caption{The background images in the (a-b) panels show cumulative NoRH ``fujiki'' radio map (distribution of brightness temperature) at 17~GHz averaged over time interval 00:00:00 - 00:39:15~UT. Red contours correspond to the cumulative NoRH ``fujiki'' radio map at 34~GHz. The panel (a) presents the cumulative AIA 304~\AA{} (blue) and 94~\AA (white) contour maps smoothed by a gaussian filter with $\sigma=1$ AIA pixel. The blue contours in the panel (b) mark levels of circular polarization at 17~GHz in percents. All contour levels are written within the panels: in units of $10^4$~K for the brightness temperature. In the panels (a) we mark some regions by denotations: the current-sheet (CS), loop-top (LT), footpoints (fp1-2), and distant footpoints (dfp1-2). The ellipses show 50~\% levels of the NoRH beams at 17 (solid) and 34~GHz (dashed) for two time instant 00:00:00 and 00:39:15~UT. The bottom panels (c-d) present simulated gyrosynchrotron $T_{br}$ spectra made for a uniform (single temperature, homogeneous) plasma with parameters written within the plots: discussion of the LT and CS microwave emission (c) and footpoints emission (d). \label{Av_MW_maps} }
\end{figure*}

Fig.\,\ref{NoRH_TPs} presents time profiles of the NoRH maximal brightness temperature (using the ``koshix'' algorithm) in the pre-flare region (a), total radio flux of the AR (c), and the circular polarization degree values -- averaged and maximal absolute value (d) in comparison with the GOES X-ray lightcurves in the both channels (b). All NoRH time profiles are made only for 17~GHz, as we have a good signal-to-noise ratio and we were not able to extract qualitative (even averaged) time profiles for 34~GHz, since they are too noisy (we see only morphology of emission sources). The time profiles in Fig.\,\ref{NoRH_TPs} show that the total radio flux (from 0.1 to 0.6 sfu) and peak temperature (from 28000 to 38000 K) are similar to the GOES time profiles by their dynamics. It points on their identical thermal plasma nature connected with the heating process around the pre-flare loop structure, where we observe the brightest emission in X-rays and in the microwave range. The averaged polarization degree is about -7~\% and the peak value is -12~\%. However using ``fujiki'' (good imaging of compact sources) we obtain larger polarization (up to 20~\%) around the pre-flare loop footpoins (Fig.\,\ref{Av_MW_maps}).

Comparison of the NoRH images with the EUV maps and time profiles clearly show that the microwave emission is generated by thermal plasma. However, we observe shifts between the regions of the maximal brightness temperature at these frequencies. We think that it may be connected to different emission mechanisms: thermal bremsstrahlung (free-free) for 34~GHz and thermal gyrosynchrotron for 17~GHz. Indeed we observe 17~GHz emission with maximal circular polarization from the pre-flare loop footpoints, where magnetic field is larger, while the brightest 34~GHz emission sources are around the loop-top (LT), where we observe the thermal X-ray source, which is produced by the thermal bremsstrahlung.

One should note that the measured brightness temperature does not exactly correspond to the real values because of the fine spatial structure within the NoRH beam. Fig.\,\ref{Av_MW_maps} presents detailed cumulative NoRH radio maps with the NoRH beams as ellipses at 50~\% levels. Even the 34~GHz beam is larger than the cross section of the LT and CS region found from the AIA EUV images. Thus, real $T_{br}$ could be a few times larger. There is also another effect leading to reducing of the temperature values in the presented cumulative radio maps in Fig.\,\ref{Av_MW_maps}: the radio map is slightly blurred by averaging effect due to the large time interval. Despite on these facts we make estimations of the observed temperatures and polarization degree (blue contours in the panel (b) using $T_{br}$ values from the current cumulative images.

To consider microwave emission at different frequencies we should take into account radiative transfer along the LOS. We are interested in two regions of the pre-flare loop: footpoints and loop-top. Let's assume the average plasma temperature of 14~MK (found from the X-ray spectra) in the regions CS, LT, and fp1, fp2 in Fig.\,\ref{Av_MW_maps}~(a). We suppose that due to thermal conduction a plasma temperature inside the magnetically coupled region is uniformly distributed. The resulted brightness temperature consists of the chromospheric part $T_{ch-br}\approx 10000$~K, coronal emission  $T_{c-br}\approx \tau_c T_c$ with $T_c\approx 3-5$ MK (typical for ARs), and emission of the compact pre-flare energy release site with a temperature of 14~MK. Let's ignore the coronal part for simplicity as $\tau_c <<1$ and we do not exactly know distribution of the coronal plasma along the LOS. If we will take brightness temperatures as $T_{34}\approx 16400$~K and $T_{17}\approx 15000$~K around the LT and CS regions (Fig.\,\ref{Av_MW_maps}), than by subtracting $T_{ch-br}$ we will estimate the real source brightness temperatures $T_{br}$: 6400 and 5000 K for 34 and 17~GHz. We think that the real $T_{br}$ for 17~GHz is higher due to effects of worse spatial resolution comparing with 34~GHz. In the case of fp1 and fp2 we can estimate the real brightness temperature as 5500~K and 20000~K for these frequencies, which can be also underestimated (possibly a few times for 17~GHz).

In  Fig.\,\ref{Av_MW_maps}~(c-d) we present models of thermal gyrosynchrotron plus bremsstrahlung microwave emission in the frequency range of 5-50~GHz for different parameters of the uniform source using the fast gyrosynchrotron codes \citep{Fleishman2010}. The first plot is about emission of the LT region. Here we take arbitrary magnetic field of 500~G and the angle between the LOS direction and magnetic field vector is equal to $70^{\circ}$ as we observe magnetic field structures near the limb mostly lying around the plane-of-sky. We consider simulated spectra for four combinations of LOS source depth (parameter $d$) and thermal plasma density $n_{th}$. Parameter $d$ is considered to be equal to 2~Mm (approximate width of the CS) and 4~Mm (the hot EUV loop-top width). The thermal plasma density in the LT and CS region is estimated from the RHESSI emission measure taken as $EM = n_{th}^2V\approx 10^{46}$~cm$^{-3}$. We consider plasma density values as $2\times 10^{10}$~cm$^{-3}$ and $4\times 10^{10}$~cm$^{-3}$ for compact volumes $V=6\times 10^{25}$~cm$^3$ and $V=1.5\times 10^{25}$~cm$^3$. Emission for all model parameters above 11~GHz is resulted from the free-free mechanism and we can obtain reasonable values of $T_br$ at 17 and 34~GHz. The best fit is for $d=2$~Mm and $n_{th}=2\times 10^{10}$~cm$^{-3}$.

To model the microwave emission around the loop footpoints fp1-2 (Fig.\,\ref{Av_MW_maps}d) we used the same values of an angle, plasma temperature and parameter $d$, but different plasma densities and magnetic field values. Plasma density is taken at lower densities of $5\times 10^9$ and $1\times10^{10}$ cm$^{-3}$ comparing with the LT region (we will discuss it further in the text). As we observe, the  strongest circular polarization and brightest emission is at 17~GHz, that's why we assume gyrosynchrotron emission in this place. We found that $B=900$~G (for comparison we also demonstrate a case with $B=1000$~G) leads to a reasonable spectra explaining brightness temperatures around 17~GHz and the selected value of plasma density explain less free-free emission at 34~GHz around the fp1-2 comparing with the LT region.


We do not model polarization values (blue contours in Fig.\,\ref{Av_MW_maps}b) in details, because of unknown distribution of the magnetic field along the LOS. Indeed due to radio waves propagation across possible quasi-transverse (QT) layers (we observe magnetic field probably around the plane-of-sky) the polarization could be modified because of the mode coupling. However, one could estimate polarization degree values for the gyrosynchrotron emission at 17~GHz using the simplified Dulk's formulas \citep{Dulk1985} and obtain real values comparable to the observed one. Generally, a polarization pattern shows a region of the optically thin plasma and there is a region with high polarization around loop footpoints (gyrosynchrotron mechanism) and smaller values at higher heights (free-free mechanism). The circular polarization degree for bremsstrahlung is estimated by a formula $r_c\approx 2 cos\Theta(\nu_B/\nu)$, where the angle between the LOS and magnetic field vector is $\Theta=70^0$ and the electron-cyclotron frequency is $\nu_B = 2.8\times10^{6B}$~Hz. Taking coronal and footpoints magnetic field values as 500 and 900~G (which were used to simulate the radio spectra shown in Fig.\,\ref{Av_MW_maps}~(c-d)), we obtain $r_c\approx 6$~\% and $r_c\approx10$~\%. Thus, the coronal polarization is explained by the free-free approach, whereas the footpoint emission should be considered in the frame of the gyrosynchrotron mechanism. The Dulk's formula (31) \citep{Dulk1985} gives a value of 45~\% for all previously used parameters of the emission source.

To sum up, we found that the microwave emission comes from the thermal plasma. The observed morphology of emission sources and the spatial distribution of circular polarization degree at 17 and 34~GHz can be explained only by different mechanisms: the free-free emission around the hot pre-flare loop-top and the presumable current sheet; the gyrosynchrotron emission around the footpoints with larger magnetic field in comparison with the higher loop-top. Despite on different emission mechanisms, the plasma temperature is fixed within the whole region, but we should consider more dense plasma around the loop-top region where we observe the pre-flare X-ray emission sources. This more dense region can be resulted from the magnetic reconnection process around the loop-top. One can consider a standing MHD shock wave leading to plasma compression. Less dense plasma filling a volume of the hot pre-flare loop is explained by absence of the intense chromospheric evaporation. Pre-flare processes are less energetic than those ones observed during the impulsive flare phase and we do not find density enhancement in the magnetic loops during the slow reconnection process inside the pre-flare current sheet. From our point of view, the case of the studied flare shows a microwave neutral line source from the side view (see more discussions in the next section). \\





{\bf 3. Results and Discussion} \\
{\bf {\it 3.1. Localization of the pre-flare current sheet}} \\

\begin{figure*}
	\centering
	\centerline{ \includegraphics[width=1.0\linewidth]{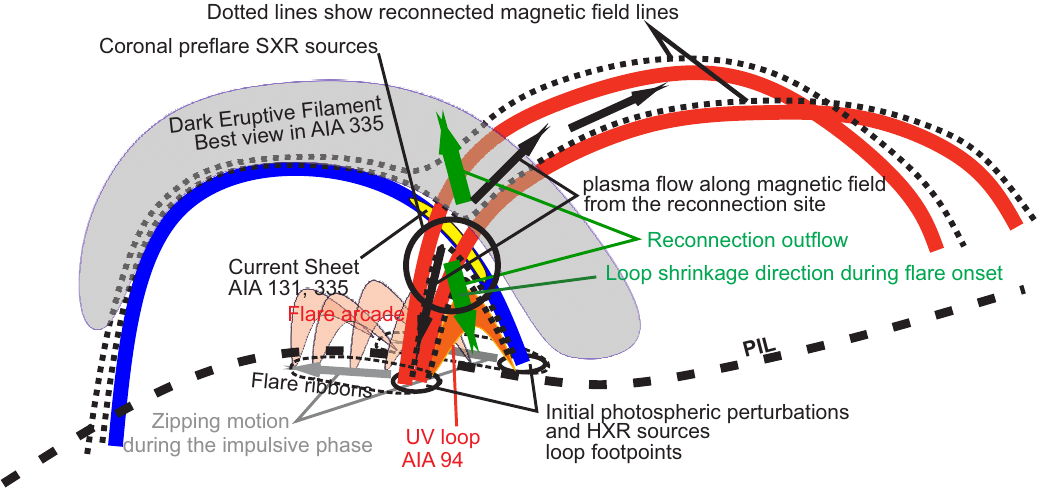} }
	\caption{Cartoon of the pre-flare and the flare onset energy release in the event studied.}
	\label{Scheme}
\end{figure*}

We summarized the main observational features of the pre-flare energy release of the studied event in the cartoon presented in Fig.\,\ref{Scheme}, which shows the magnetic field geometry in the frames of the TCMR scenario. One of the key results of this research is localization of the pre-flare current sheet relatively to the future initial flare energy release site and the eruptive structure. From our point of view it is the first observation of the pre-flare magnetic reconnection studied simultaneously in the wide range of wavelengths (in microwaves, EUV, and X-rays). We established a set of implicit evidences of the TCMR pre-flare current sheet based on the following observational peculiarities and specific morphological features for the studied solar flare event:

\begin{enumerate}
	\item The observed (intersecting in the plane-of-sky) large-scale EUV loop-like structures tracing magnetic field geometry are favorable for the current sheet formation in the frame of the TCMR scenario. It was also found and discussed in the work of \cite{Chen2014};
	\item Distribution of the pre-flare hot EUV and X-ray sources are likely due to the energy release in the assumed current sheet, which is formed in the intersecting large-scale coronal magnetic loops. We observe a possible place of the current sheet as the thin EUV source between two magnetic structures associated with the magnetic reconnection during the pre-flare phase -- the hot loop below the current sheet and the dark eruptive structure above the current sheet;
	\item The previous statement is also confirmed by the DEM analysis, which shows the thin elongated region (sometimes fragmented) with the enhanced temperatures (up to 20~MK) possibly associated with the current sheet in the place of interacting sheared magnetic loops;
	\item The performed spectral analysis of X-rays and DEM reconstruction confirms an appearance of the high temperature (up to 20~MK) plasma, which can be connected with the magnetic reconnection process and corresponding intensive heating. This plasma is definitely much hotter than plasmas in the surrounding medium;
	\item The structure of the plasma flows in the corona can be connected with the reconnecting process inside the pre-flare current sheet. We observed plasma flows in the ``hot'' AIA channels (coronal temperatures) around the current sheet and in the cold 304~\AA{} channel. It is worth noting that the dark filament is likely became larger due to the pre-flare reconnection process and flows directed from the current sheet to the filament body \citep[TCMR model of ][]{Moore2001};
	\item The downward flare loop shrinkage and the upward initial motion of the eruptive structure during the flare onset was started from the position of the assumed pre-flare current sheet. Thus, one can conclude that the place of the initial flare energy release is co-spatial and directly connected with the pre-flare energy release site. We see transition from the slow magnetic reconnection to the flare fast regime.
	\item We also found signatures of the initial flare energy release in the place, which is assumed to be the pre-flare current sheet -- an appearance of the EUV brightenings and their motions and a subsequent start of the eruption and the flare onset; 
	\item Appearance of the double HXR sources and photospheric perturbations in the shrinking loop around the flare onset time. The flare high-energy release is also connected with the hot pre-flare loop lying below the pre-flare current sheet;
	\item The main photospheric magnetic field change was in the place between the photospheric perturbations and the flare HXR sources below the shrinking loop (during the flare onset). This magnetic field variations are likely due to magnetic reconnection in the current sheet and subsequent magnetic field advection seen as the loop shrinkage;
	\item Analysis of the microwave NoRH images at 17 and 34~GHz has shown that the loop-top (probably above the loop-top) emission is explained by denser plasma than emission of the loop itself. It could be explained by plasma compression due to pile-up of reconnected magnetic flux tubes came from the pre-flare current sheet.
\end{enumerate}

Fig.\,\ref{Scheme} shows an interacting magnetic tube (blue) with a bunch of other red ones. As a result, we have a thin region of the possible current sheet localization in this three-dimensional magnetic configuration. One can also consider another geometric concepts, but for us it seems to be a realistic one as we have evidences of such scenario in the AIA EUV images. One can suspect a few possible simple 3D geometric schemes of interacting sheared magnetic structures in the frame of the TCMR-like geometry in the cases of other flares: loop-to-loop (point like intersection), loop-to-bunch (thin region, as in our case), bunch-to-bunch (2D plane region) and irregular (many loops and bunches). To study these possible geometries one should conduct many other case-studies and statistical researches.

Multiwavelength analysis of energies shows that a dominating energy release channel is a changing internal plasma energy in the wide temperature range from a few MK up to 20~MK. The maximal heating rate $\approx 2\times 10^{27}$~erg/s is revealed by the DEM analysis. Some heating events according to the DEM are characterized by a heating power of $\approx 5\times 10^{26}$~erg/s. Using the characteristic length scale of the current sheet $L \approx 13$~Mm (estimated from the EUV images), magnetic field $B\approx 500$~G (rough estimations from the microwave data, see discussions in the previous section) and plasma density $n \sim 10^{10}$~cm$^{-3}$ (based on the EM from the X-ray spectra) one can estimate an energy release rate in the current sheet: $d\epsilon_m/dt\approx 10^{22}(B/100~G)^2(L/1~Mm)^2(v_a/1~Mm/s)\approx 4\times 10^{26}$~erg/s for $v_a\approx 10$~Mm/s. Thus, the energy release rate in the quasi-stationary regime can explain the observed total heating power in the pre-flare phase.

However, the X-ray spectral analysis showed that the presence of nonthermal electrons was possible and their energetics could be comparable with the radiative losses and even with the time derivative of the X-ray emitting plasma internal energy.

We should also discuss morphology of the X-ray and EUV sources around the assumed current sheet. There were three features: the hot loop, the thin EUV source, and the bright X-ray emitting region around the loop-top (there is also the brightest microwave 34~GHz emission source there). We suppose that the bright loop is a manifestation of the reconnected magnetic field lines: possibly, plasma heating of the loop-top region could be due to stationary shock waves. The thin EUV source is assumed to be around the current sheet as only here we found plasma temperatures up to 20~MK according to the DEM analysis, when the X-ray analysis revealed only the temperatures in the range of 9-14~MK (with one heating event up to 17~MK) around the loop-top.

The EUV images point on the fine spatial structure of the assumed current sheet in a form of small-scale bright blobs forming a chainy EUV thin source. This peculiarity is best seen in the DEM maps at temperatures above 15~MK indicating small-scale energy release episodes. We do not know the nature of such irregularity but we can assume separate reconnection events of the intersecting magnetic thin loops or tearing instability of the current sheet \citep{Priest2000_book}. Spatial modulation of energy release inside the current sheet is also possible due to some waves in the magnetized plasma \citep[e.g., ][]{ArtemyevZimovets2012} and thermal instability \citep[][]{Ledentsov2021a,Ledentsov2021b,Ledentsov2021c}. \\




{\bf {\it 3.2. Formation of the magnetic flux rope and eruption onset}} \\

One of the central questions of the solar physics is how magnetic flux ropes form and erupted that lead to CMEs. This problem was considered in multitude of works both by theoretical and observational techniques. In the case of this work, we found an exceptional solar flare which is suitable for very detailed multiwavelength analysis of the pre-flare processes and the transition to the flare impulsive phase and progressing eruption. There are two types of instabilities, which can lead to eruption: ideal or non-ideal/resistive instabilities.

CMEs are eruptions of sheared or/and twisted magnetic structures commonly called magnetic flux ropes. They are observed remotely in the solar atmosphere and detected in situ in the interplanetary space. Understanding what initiates and drives solar eruptions belongs to the highest priority goals in the solar and heliospheric physics. The currently debated categories of ideal MHD instabilities and reconnection models under the energy storage-and-release paradigm correspond, respectively, to the assumptions of a flux rope or sheared arcade as the magnetic structure at the onset of an eruption. Although it has been established that a flux rope is ultimately created, it remains yet to be clarified whether the onset is dominated by the torus instability \citep[][]{vanTend1978, ForbesIsenberg1991, KliemTorok2006, IshiguroKusano2017}, magnetic reconnection \citep[][]{Moore2001, Karpen2012, Kusano2012}, or a combination of both.

The theoretical onset condition is only available for the torus instability yet, although several different drivers and evolutionary sequences can be involved. The magnetic and thermal structures at the onset of the eruption likely require certain key properties, e.g., a certain ratio of the flux in the filament channel vs. the active-region flux \citep[][]{Su2011, Savcheva2012a}, or a minimum value of a helicity measure \citep[e.g. ][]{Zuccarello2018}, or a sufficiently coherent flux rope with a hyperbolic flux tube underneath \citep[][]{Guo2010, Aulanier2010, Savcheva2012b}. Magnetic reconnection can lead to loss of stability of a twisted and prepared for eruption magnetic flux rope \citep[e.g. ][]{ForbesPriest1995}.

Analysis of the EUV images in different AIA channels in this work allowed us to find connections between the eruptive filament growth and activity of the pre-flare current sheet. We found that the magnetic flux rope was formed above the current sheet location due to magnetic reconnection via the TCMR process \citep{Moore2001}. We do not deny a role of ideal MHD effects. Indeed a helical magnetic structure has a tendency to experience more tensions due to magnetic forces: more twisted magnetic structure leads to more probable eruption onset (depending on surrounding magnetized medium). The question is: how twist is generated? In our case a twist number is not estimated because we do not have possibility to make NLFFF reconstruction of the coronal magnetic field (because of the near the limb position). However, one can suspect that the twist is more and more enhanced due to continuous inflow of the poloidal magnetic field coming from the pre-flare magnetic reconnection region. At some time instant, a single reconnection episode (trigger) finally could destabilize a preliminarily twisted (ready for eruption) overall magnetic structure and the flare started its energy release. In other words, the studied case clearly shows the resistive MHD nature (in contrast to the pure ideal MHD scenario of the eruptive flare) of the pre-flare and the pre-eruptive state of the large eruptive two-ribbon solar flare. Moreover, a small microflare is likely a final trigger of overall energy release. \\

{\bf {\it 3.3. Pre-flare and flare onset timing}} \\

\begin{figure*}
	\centering
	\centerline{ \includegraphics[width=1.0\linewidth]{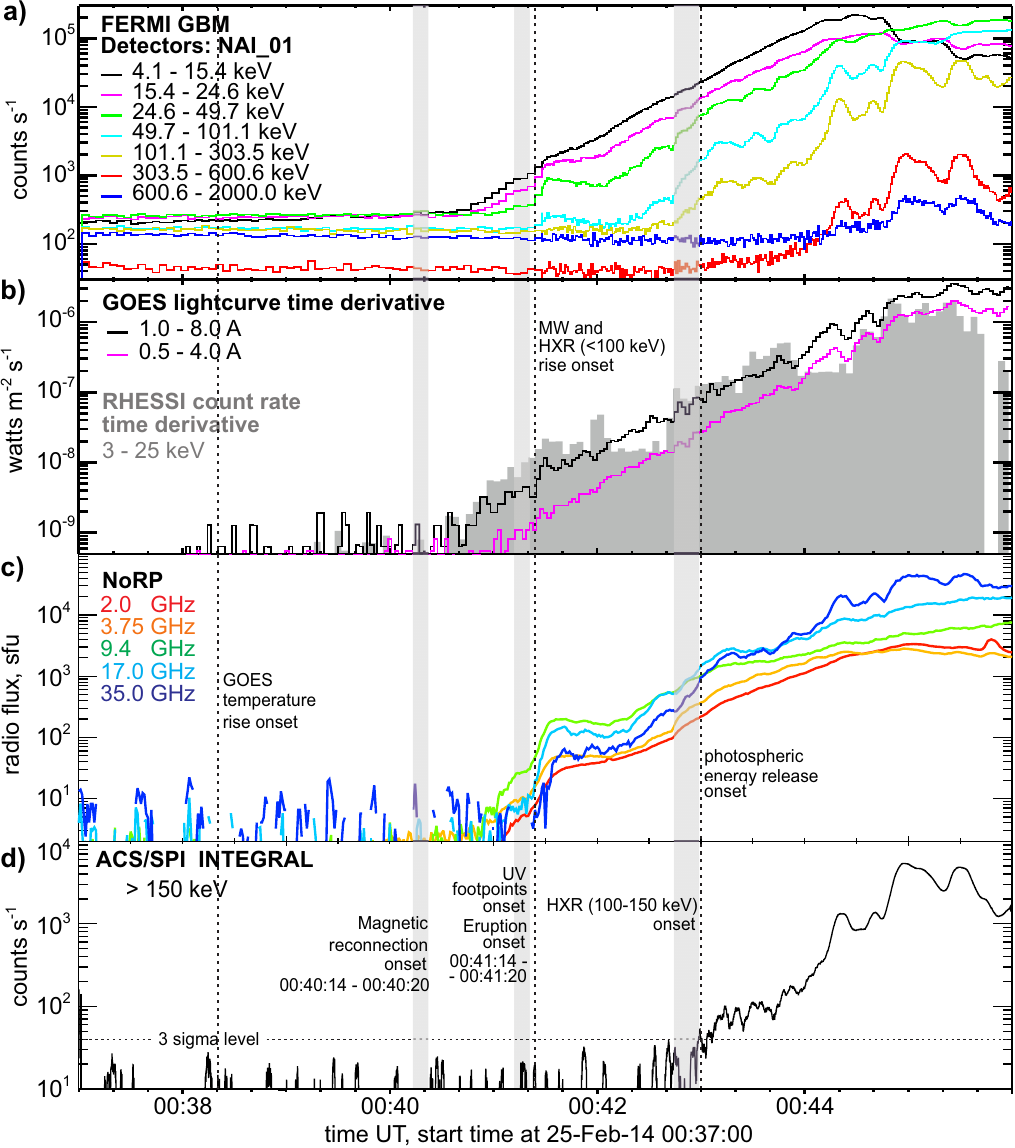} }
	\caption{Overview of the X-ray and microwave time profiles for the time period covering a part of the pre-flare phase, the flare onset and a part of the impulsive phase. The vertical dashed lines and gray stripes show the time instants of the different phenomena occurred in the pre-flare and flare region.}
	\label{TPs_aroundOnset}
\end{figure*}

In Fig.\,\ref{TPs_aroundOnset} we present the flare ``history'' around its onset: a summary of the temporal profiles from the FERMI GBM NaI\_01 4-2000~keV (a), GOES two lightcurve time derivatives with the RHESSI count rate 3-25 keV (b), NoRP microwave flux 2-35 GHz (c), and X-ray ACS/SPI INTEGRAL count rate above at energies above $\approx$100~keV (d). We argue that in the frame of the performed research, we reconstructed a detailed energy release chronology from the pre-flare time period to the initial time period of the impulsive phase. We have a few important time stamps after the filament growth period: the pre-heating onset (start of the GOES temperature rise), trigger in a form of the magnetic reconnection onset, eruption onset and appearance of the HXR emission approximately above 100~keV. The longest time period is the pre-heating stage, which was about 2~min. The time delay between the trigger and eruption onset was about 1~min. Possibly, the pre-heating time period is explained (by order-of-magnitude estimations) by heating conduction time $\tau_c = 4.2k_Bn_eL^2/(kT_e^{5/2})\approx 40$~s, where $k=9.2\times 10^{-7}$~erg\,s$^{-1}$cm$^{-1}$K$^{-7/2}$ is the thermal Spitzer conductivity coefficient, the length scale $L\approx 10$~Mm, plasma density $n_e\approx 2\times 10^{10}$~cm$^{-3}$, and plasma temperature $T_e\approx 10$~MK.

The start of the eruption onset is possibly delayed due to time of the tearing mode development $\tau_{tear}\sim \sqrt{\tau_d\tau_A}$ (mean geometric of the magnetic diffusion and Alfven crossing times) for the longest wavelength perturbations of the current sheet with a width $l$. This instability is connected with the finite plasma resistivity. Considering $\tau_{tear}\sim 60$~s and $\tau_A\sim 1$~s we should consider magnetic diffusivity much larger comparing with the classic values, which are limited only by collisions. Detailed discussions of the plasma turbulence inside and around the current sheet are impossible in the frame of the used data.

Appearance of the energetic emissions above 100~keV (around 90~s after the eruption onset) is connected with the eruption process, which forms the flare current sheets during the impulsive phase. Details of the energy release in the impulsive phase are described in the work of \cite{Sharykin2023}, where we discuss two stages of magnetic reconnection. The initial flare energy release started in the sheared magnetic loops around the PIL and then we observed a classical situation of the current sheet below the moving eruptive magnetic flux rope. \\

{\bf {\it 3.4. The observed microwave sources as a side-view on NLS}} \\

In this paper we considered the morphology of the pre-flare microwave sources seen at 17 and 34~GHz in details from the side view considering the near limb location of the flare. We think that the observed distribution of the microwave brightness temperature and polarization at 17~GHz is in favor of the  presence of the rather compact Neutral-Line Source \citep[NLS, e.g.][]{Uralov2006, Uralov2008} embedded into the relatively larger diffuse emission region, but we see it from the side view. Previously, NLSs were studied from the point of view of the magnetic field structure and there were no detailed research using X-ray and EUV maps, which could demonstrate morphology of the energy release site. Thus, in the frame of the current work, we retrieved a rich multiwavelength information about the processes around the NLS close to the flare and found that the observed NLS is of thermal origin and connected with the pre-flare magnetic reconnection in the frame of the TCMR process. Moreover, we can state that all (diffuse and compact) microwave sources are connected via large-scale magnetic loops interacting around the probable pre-flare current sheet.

The found connections of the X-ray and EUV precursors (which are connected with the flare and eruption onset) with the NLS show perspective of the flare prediction using observations of NLSs and investigation of its dynamics. It has been previously shown that radio observations can be used to predict proton flares \citep[e.g. ][]{TanakaEnome1975, Bogod2018, Kim2018} and flares in general \citep[e.g. ][]{Smolkov2010, Shramko2019}. Further investigations of the pre-flare NLS peculiarities can help to improve prediction methods, because microwave observations are sensitive both for high temperature thermal plasma and magnetic field where plasma is confined. \\

{\bf 4. Conclusions} \\

In this research, we do not propose new concepts or discover new features during the pre-flare process, but consider well-known pre-flare and flare phenomena. Our methodological peculiarity is to trace energy release from the pre-flare state to the beginning of the impulsive phase continuously (for around 45~min). This research summarizes a large data set for the pre-flare dynamics before an X-class solar flare and, to our mind, results in the deeper insight in the tether-cutting process.

We reconstructed a detailed chronology of the pre-flare energy release and its transition to the impulsive phase for the case of a powerful eruptive X4.9-class solar flare on February 25, 2014 using multiwavelength data (ultraviolet, microwave, X-ray and visible emissions). Possibly for the first time, we were able to localize the pre-flare current sheet region in the tether-cutting geometry using a set of indirect features, and obtained the following important conclusions concerning the preparation of the AR to a powerful flare, eruption and CME and its initiation:

\begin{enumerate}

	\item The pre-flare energy region was localized by using simultaneous multiwavelength observations. This region is quite compact ($<$10~Mm) and appeared as a result of the interaction (the tether-cutting type) of large-scale ($\sim$50~Mm) magnetic structures near the PIL. An analysis of the different types of emissions, the morphology of the EUV/UV sources, the observed plasma flows, and its dynamics around the flare onset indicate that a pre-flare current sheet was located in this region. From our point of view, this is one of the first indirect localizations and detailed multiwavelength study of the region of the pre-flare current sheet. We proved causal relationship between the pre-flare energy release and the flare/eruption onset seen in the various ranges of the electromagnetic spectrum;
	
	\item We showed the relationship between the energy release dynamics in the region of the pre-flare current layer and the formation of the eruptive magnetic flux rope. The growth and slow rise of the magnetic flux rope (prominence) was associated with the activation of local plasma emissions and increase in UV radiation fluxes from the region where the current layer is located. The prominence gradually increased its mass due to the inflows of plasma with magnetic field lines ejected from the reconnecting pre-flare current layer. The obtained observational results confirm the tether-cutting model, which assumes the formation of an eruptive flux rope in the pre-flare phase due to magnetic reconnection in the (non-neutral) current layer near the PIL;
	
	\item We showed that the most probable trigger of the eruption is a local fast magnetic reconnection episode of the ``microflare'' type (in terms of its energy) in the pre-flare current layer. It is worth explaining that some local instability in the pre-flare current layer results in a transition from the slow to the fast regime of magnetic reconnection, which initiates ejection from the current layer and disrupts the equilibrium of the eruptive magnetic flux rope (which was probably prepared for eruption, e.g. twisted). At the same time, the erupting flux rope itself initiates the formation of the main reconnecting flare current layer (below the flux rope) during its rise, which is described by the ``standard'' model of an eruptive two-ribbon flare. The appearance of the photospheric sources of optical radiation and energetic hard X-rays, as well as type III radio bursts, are already associated with the development of an eruptive current layer in the impulsive phase \citep[for details see][]{Sharykin2023};
	
	\item We obtained unique detailed multiwavelength information about the microwave neutral-line source (NLS) from the side view before the large flare and argued that this microwave source was due to magnetic reconnection inside the pre-flare current sheet in the frame of the tether-cutting scenario.
	
\end{enumerate}

This work clearly shows connection between the pre-flare processes and the flare initiation and start of the eruption. Probably, having more powerful (better spatial and temporal resolutions) observational instruments, we shall typically observe emission sources in X-rays, ultraviolet, and microwave ranges connected with the pre-flare current sheets. Collecting statistics of pre-flare energy release (precursors) at different intensity amplitudes will give additional information about preparation of ARs to flares. We believe that future empirical solar flare forecast techniques will use not only photospheric magnetograms but also emission maps providing us with the current and recent energy release status of ARs (an AR's ``history'' of energy release).


{\bf Acknowledgements}
This work is supported by the Russian Science Foundation under grant 20-72-10158.

\bibliographystyle{plain}

\begin{thebibliography}{10}

\bibitem{AbramovMaximov2015a}
V.~E. {Abramov-Maximov}, V.~N. {Borovik}, L.~V. {Opeikina}, and A.~G. {Tlatov}.
\newblock {Dynamics of Microwave Sources Associated with the Neutral Line and
  the Magnetic-Field Parameters of Sunspots as a Factor in Predicting Large
  Flares}.
\newblock {\em \solphys}, 290(1):53--77, January 2015.

\bibitem{ArtemyevZimovets2012}
A.~{Artemyev} and I.~{Zimovets}.
\newblock {Stability of Current Sheets in the Solar Corona}.
\newblock {\em \solphys}, 277(2):283--298, April 2012.

\bibitem{Aschwanden2015}
Markus~J. {Aschwanden}, Paul {Boerner}, Daniel {Ryan}, Amir {Caspi}, James~M.
  {McTiernan}, and Harry~P. {Warren}.
\newblock {Global Energetics of Solar Flares: II. Thermal Energies}.
\newblock {\em \apj}, 802(1):53, March 2015.

\bibitem{Aulanier2010}
G.~{Aulanier}, T.~{T{\"o}r{\"o}k}, P.~{D{\'e}moulin}, and E.~E. {DeLuca}.
\newblock {Formation of Torus-Unstable Flux Ropes and Electric Currents in
  Erupting Sigmoids}.
\newblock {\em \apj}, 708(1):314--333, January 2010.

\bibitem{Bakunina2015}
I.~A. {Bakunina}, V.~F. {Melnikov}, A.~A. {Solov'ev}, and V.~E.
  {Abramov-Maximov}.
\newblock {Intersunspot Microwave Sources}.
\newblock {\em \solphys}, 290(1):37--52, January 2015.

\bibitem{Bogod2018}
V.~M. {Bogod}, P.~M. {Svidskiy}, E.~A. {Kurochkin}, A.~V. {Shendrik}, and N.~P.
  {Everstov}.
\newblock {A Method of Forecasting Solar Activity Based on Radio Astronomical
  Observations}.
\newblock {\em Astrophysical Bulletin}, 73(4):478--486, October 2018.

\bibitem{Brown1971}
J.~C. {Brown}.
\newblock {The Deduction of Energy Spectra of Non-Thermal Electrons in Flares
  from the Observed Dynamic Spectra of Hard X-Ray Bursts}.
\newblock {\em \solphys}, 18:489--502, July 1971.

\bibitem{Chen2014}
Huadong {Chen}, Jun {Zhang}, Xin {Cheng}, Suli {Ma}, Shuhong {Yang}, and Ting
  {Li}.
\newblock {Direct Observations of Tether-cutting Reconnection during a Major
  Solar Event from 2014 February 24 to 25}.
\newblock {\em \apjl}, 797(2):L15, December 2014.

\bibitem{Chifor2006}
C.~{Chifor}, H.~E. {Mason}, D.~{Tripathi}, H.~{Isobe}, and A.~{Asai}.
\newblock {The early phases of a solar prominence eruption and associated
  flare: a multi-wavelength analysis}.
\newblock {\em \aap}, 458(3):965--973, November 2006.

\bibitem{Chifor2007}
C.~{Chifor}, D.~{Tripathi}, H.~E. {Mason}, and B.~R. {Dennis}.
\newblock {X-ray precursors to flares and filament eruptions}.
\newblock {\em \aap}, 472(3):967--979, September 2007.

\bibitem{Dulk1985}
G.~A. {Dulk}.
\newblock {Radio emission from the sun and stars.}
\newblock {\em \araa}, 23:169--224, January 1985.

\bibitem{Fleishman2010}
Gregory~D. {Fleishman} and Alexey~A. {Kuznetsov}.
\newblock {Fast Gyrosynchrotron Codes}.
\newblock {\em \apj}, 721(2):1127--1141, October 2010.

\bibitem{ForbesIsenberg1991}
T.~G. {Forbes} and P.~A. {Isenberg}.
\newblock {A Catastrophe Mechanism for Coronal Mass Ejections}.
\newblock {\em \apj}, 373:294, May 1991.

\bibitem{ForbesPriest1995}
T.~G. {Forbes} and E.~R. {Priest}.
\newblock {Photospheric Magnetic Field Evolution and Eruptive Flares}.
\newblock {\em \apj}, 446:377, June 1995.

\bibitem{Gibson2006}
S.~E. {Gibson}, Y.~{Fan}, T.~{T{\"o}r{\"o}k}, and B.~{Kliem}.
\newblock {The Evolving Sigmoid: Evidence for Magnetic Flux Ropes in the Corona
  Before, During, and After CMES}.
\newblock {\em \ssr}, 124(1-4):131--144, June 2006.

\bibitem{Guo2010}
Y.~{Guo}, M.~D. {Ding}, B.~{Schmieder}, H.~{Li}, T.~{T{\"o}r{\"o}k}, and
  T.~{Wiegelmann}.
\newblock {Driving Mechanism and Onset Condition of a Confined Eruption}.
\newblock {\em \apjl}, 725(1):L38--L42, December 2010.

\bibitem{Hannah2012}
I.~G. {Hannah} and E.~P. {Kontar}.
\newblock {Differential emission measures from the regularized inversion of
  Hinode and SDO data}.
\newblock {\em \aap}, 539:A146, March 2012.

\bibitem{Hannah2013}
I.~G. {Hannah} and E.~P. {Kontar}.
\newblock {Multi-thermal dynamics and energetics of a coronal mass ejection in
  the low solar atmosphere}.
\newblock {\em \aap}, 553:A10, May 2013.

\bibitem{Hirayama1974}
T.~{Hirayama}.
\newblock {Theoretical Model of Flares and Prominences. I: Evaporating Flare
  Model}.
\newblock {\em \solphys}, 34(2):323--338, February 1974.

\bibitem{IshiguroKusano2017}
N.~{Ishiguro} and K.~{Kusano}.
\newblock {Double Arc Instability in the Solar Corona}.
\newblock {\em \apj}, 843(2):101, July 2017.

\bibitem{Karpen2012}
J.~T. {Karpen}, S.~K. {Antiochos}, and C.~R. {DeVore}.
\newblock {The Mechanisms for the Onset and Explosive Eruption of Coronal Mass
  Ejections and Eruptive Flares}.
\newblock {\em \apj}, 760(1):81, November 2012.

\bibitem{Kim2018}
Kyong~Nam {Kim}, Sun~Ae {Sin}, Kum~Ae {Song}, and Jin~Hyok {Kong}.
\newblock {A technique for prediction of SPEs from solar radio flux by
  statistical analysis, ANN and GA}.
\newblock {\em \apss}, 363(8):170, August 2018.

\bibitem{KliemTorok2006}
B.~{Kliem} and T.~{T{\"o}r{\"o}k}.
\newblock {Torus Instability}.
\newblock {\em \prl}, 96(25):255002, June 2006.

\bibitem{Kudriavtseva2021}
Anastasiia {Kudriavtseva}, Ivan {Myshyakov}, Arkadiy {Uralov}, and Victor
  {Grechnev}.
\newblock {Microwave indicator of potential geoeffectiveness and magnetic
  flux-rope structure of a solar active region}.
\newblock {\em Solar-Terrestrial Physics}, 7(1):3--10, March 2021.

\bibitem{Kusano2012}
K.~{Kusano}, Y.~{Bamba}, T.~T. {Yamamoto}, Y.~{Iida}, S.~{Toriumi}, and
  A.~{Asai}.
\newblock {Magnetic Field Structures Triggering Solar Flares and Coronal Mass
  Ejections}.
\newblock {\em \apj}, 760(1):31, November 2012.

\bibitem{Ledentsov2021a}
Leonid {Ledentsov}.
\newblock {Thermal Trigger for Solar Flares I: Fragmentation of the Preflare
  Current Layer}.
\newblock {\em \solphys}, 296(4):74, April 2021.

\bibitem{Ledentsov2021b}
Leonid {Ledentsov}.
\newblock {Thermal Trigger for Solar Flares II: Effect of the Guide Magnetic
  Field}.
\newblock {\em \solphys}, 296(6):93, June 2021.

\bibitem{Ledentsov2021c}
Leonid {Ledentsov}.
\newblock {Thermal Trigger for Solar Flares III: Effect of the Oblique Layer
  Fragmentation}.
\newblock {\em \solphys}, 296(8):117, August 2021.

\bibitem{Lemen2012}
J.~R. {Lemen}, A.~M. {Title}, D.~J. {Akin}, P.~F. {Boerner}, C.~{Chou}, J.~F.
  {Drake}, D.~W. {Duncan}, C.~G. {Edwards}, F.~M. {Friedlaender}, G.~F.
  {Heyman}, N.~E. {Hurlburt}, N.~L. {Katz}, G.~D. {Kushner}, M.~{Levay}, R.~W.
  {Lindgren}, D.~P. {Mathur}, E.~L. {McFeaters}, S.~{Mitchell}, R.~A. {Rehse},
  C.~J. {Schrijver}, L.~A. {Springer}, R.~A. {Stern}, T.~D. {Tarbell}, J.-P.
  {Wuelser}, C.~J. {Wolfson}, C.~{Yanari}, J.~A. {Bookbinder}, P.~N.
  {Cheimets}, D.~{Caldwell}, E.~E. {Deluca}, R.~{Gates}, L.~{Golub}, S.~{Park},
  W.~A. {Podgorski}, R.~I. {Bush}, P.~H. {Scherrer}, M.~A. {Gummin},
  P.~{Smith}, G.~{Auker}, P.~{Jerram}, P.~{Pool}, R.~{Soufli}, D.~L. {Windt},
  S.~{Beardsley}, M.~{Clapp}, J.~{Lang}, and N.~{Waltham}.
\newblock {The Atmospheric Imaging Assembly (AIA) on the Solar Dynamics
  Observatory (SDO)}.
\newblock {\em \solphys}, 275:17--40, January 2012.

\bibitem{LiGan2005}
Y.~P. {Li} and W.~Q. {Gan}.
\newblock {The Shrinkage of Flare Radio Loops}.
\newblock {\em \apjl}, 629(2):L137--L139, August 2005.

\bibitem{Lin2002}
R.~P. {Lin}, B.~R. {Dennis}, G.~J. {Hurford}, D.~M. {Smith}, A.~{Zehnder},
  P.~R. {Harvey}, D.~W. {Curtis}, D.~{Pankow}, P.~{Turin}, M.~{Bester},
  A.~{Csillaghy}, M.~{Lewis}, N.~{Madden}, H.~F. {van Beek}, M.~{Appleby},
  T.~{Raudorf}, J.~{McTiernan}, R.~{Ramaty}, E.~{Schmahl}, R.~{Schwartz},
  S.~{Krucker}, R.~{Abiad}, T.~{Quinn}, P.~{Berg}, M.~{Hashii}, R.~{Sterling},
  R.~{Jackson}, R.~{Pratt}, R.~D. {Campbell}, D.~{Malone}, D.~{Landis}, C.~P.
  {Barrington-Leigh}, S.~{Slassi-Sennou}, C.~{Cork}, D.~{Clark}, D.~{Amato},
  L.~{Orwig}, R.~{Boyle}, I.~S. {Banks}, K.~{Shirey}, A.~K. {Tolbert},
  D.~{Zarro}, F.~{Snow}, K.~{Thomsen}, R.~{Henneck}, A.~{McHedlishvili},
  P.~{Ming}, M.~{Fivian}, J.~{Jordan}, R.~{Wanner}, J.~{Crubb}, J.~{Preble},
  M.~{Matranga}, A.~{Benz}, H.~{Hudson}, R.~C. {Canfield}, G.~D. {Holman},
  C.~{Crannell}, T.~{Kosugi}, A.~G. {Emslie}, N.~{Vilmer}, J.~C. {Brown},
  C.~{Johns-Krull}, M.~{Aschwanden}, T.~{Metcalf}, and A.~{Conway}.
\newblock {The Reuven Ramaty High-Energy Solar Spectroscopic Imager (RHESSI)}.
\newblock {\em \solphys}, 210:3--32, November 2002.

\bibitem{Liu2013}
Chang {Liu}, Na~{Deng}, Jeongwoo {Lee}, Thomas {Wiegelmann}, Ronald~L. {Moore},
  and Haimin {Wang}.
\newblock {Evidence for Solar Tether-cutting Magnetic Reconnection from Coronal
  Field Extrapolations}.
\newblock {\em \apjl}, 778(2):L36, December 2013.

\bibitem{Longcope2018}
Dana {Longcope}, John {Unverferth}, Courtney {Klein}, Marika {McCarthy}, and
  Eric {Priest}.
\newblock {Evidence for Downflows in the Narrow Plasma Sheet of 2017 September
  10 and Their Significance for Flare Reconnection}.
\newblock {\em \apj}, 868(2):148, December 2018.

\bibitem{Magara1996}
Tetsuya {Magara}, Shin {Mineshige}, Takaaki {Yokoyama}, and Kazunari {Shibata}.
\newblock {Numerical Simulation of Magnetic Reconnection in Eruptive Flares}.
\newblock {\em \apj}, 466:1054, August 1996.

\bibitem{Moore2001}
Ronald~L. {Moore}, Alphonse~C. {Sterling}, Hugh~S. {Hudson}, and James~R.
  {Lemen}.
\newblock {Onset of the Magnetic Explosion in Solar Flares and Coronal Mass
  Ejections}.
\newblock {\em \apj}, 552(2):833--848, May 2001.

\bibitem{Pesnell2012}
W.~D. {Pesnell}, B.~J. {Thompson}, and P.~C. {Chamberlin}.
\newblock {The Solar Dynamics Observatory (SDO)}.
\newblock {\em \solphys}, 275:3--15, January 2012.

\bibitem{Priest2000_book}
Eric {Priest} and Terry {Forbes}.
\newblock {\em {Magnetic Reconnection: MHD Theory and Applications}}.
\newblock 2000.

\bibitem{Ryan2012}
Daniel~F. {Ryan}, Ryan~O. {Milligan}, Peter~T. {Gallagher}, Brian~R. {Dennis},
  A.~Kim {Tolbert}, Richard~A. {Schwartz}, and C.~Alex {Young}.
\newblock {The Thermal Properties of Solar Flares over Three Solar Cycles Using
  GOES X-Ray Observations}.
\newblock {\em \apjs}, 202(2):11, October 2012.

\bibitem{Savcheva2012b}
A.~{Savcheva}, E.~{Pariat}, A.~{van Ballegooijen}, G.~{Aulanier}, and
  E.~{DeLuca}.
\newblock {Sigmoidal Active Region on the Sun: Comparison of a
  Magnetohydrodynamical Simulation and a Nonlinear Force-free Field Model}.
\newblock {\em \apj}, 750(1):15, May 2012.

\bibitem{Savcheva2012a}
A.~S. {Savcheva}, L.~M. {Green}, A.~A. {van Ballegooijen}, and E.~E. {DeLuca}.
\newblock {Photospheric Flux Cancellation and the Build-up of Sigmoidal Flux
  Ropes on the Sun}.
\newblock {\em \apj}, 759(2):105, November 2012.

\bibitem{Scherrer2012}
P.~H. {Scherrer}, J.~{Schou}, R.~I. {Bush}, A.~G. {Kosovichev}, R.~S. {Bogart},
  J.~T. {Hoeksema}, Y.~{Liu}, T.~L. {Duvall}, J.~{Zhao}, A.~M. {Title}, C.~J.
  {Schrijver}, T.~D. {Tarbell}, and S.~{Tomczyk}.
\newblock {The Helioseismic and Magnetic Imager (HMI) Investigation for the
  Solar Dynamics Observatory (SDO)}.
\newblock {\em \solphys}, 275:207--227, January 2012.

\bibitem{Sharykin2017}
I.~N. {Sharykin}, A.~G. {Kosovichev}, V.~M. {Sadykov}, I.~V. {Zimovets}, and
  I.~I. {Myshyakov}.
\newblock {Investigation of Relationship between High-energy X-Ray Sources and
  Photospheric and Helioseismic Impacts of X1.8 Solar Flare of 2012 October
  23}.
\newblock {\em \apj}, 843(1):67, July 2017.

\bibitem{Sharykin2015c}
I.~N. {Sharykin}, A.~B. {Struminskii}, and I.~V. {Zimovets}.
\newblock {Plasma heating to super-hot temperatures (>30 MK) in the August 9,
  2011 solar flare}.
\newblock {\em Astronomy Letters}, 41(1-2):53--66, January 2015.

\bibitem{Sharykin2020}
I.~N. {Sharykin}, I.~V. {Zimovets}, and I.~I. {Myshyakov}.
\newblock {Flare Energy Release at the Magnetic Field Polarity Inversion Line
  during the M1.2 Solar Flare of 2015 March 15. II. Investigation of
  Photospheric Electric Current and Magnetic Field Variations Using HMI 135 s
  Vector Magnetograms}.
\newblock {\em \apj}, 893(2):159, April 2020.

\bibitem{Sharykin2018}
I.~N. {Sharykin}, I.~V. {Zimovets}, I.~I. {Myshyakov}, and N.~S. {Meshalkina}.
\newblock {Flare Energy Release at the Magnetic Field Polarity Inversion Line
  during the M1.2 Solar Flare of 2015 March 15. I. Onset of Plasma Heating and
  Electron Acceleration}.
\newblock {\em \apj}, 864(2):156, September 2018.

\bibitem{Sharykin2023}
I.~N. {Sharykin}, I.~V. {Zimovets}, and A.~V. {Radivon}.
\newblock {High-Cadence Observations of Magnetic Field Dynamics and
  Photospheric Emission Sources in the Eruptive Near-the-Limb X4.9 Solar Flare
  on 25 February, 2014: Evidences for Two-Stage Magnetic Reconnection during
  the Impulsive Phase}.
\newblock {\em Cosmic Research}, 61(4):265--282, August 2023.

\bibitem{Shramko2019}
A.~D. {Shramko}, A.~G. {Tlatov}, and S.~A. {Guseva}.
\newblock {Forecast of solar flares from observations in the microwave range at
  the Kislovodsk station}.
\newblock {\em Astronomical and Astrophysical Transactions}, 31(2):103--108,
  January 2019.

\bibitem{Smolkov2010}
G.~Ya. {SmolKov}, V.~P. {Maksimov}, D.~V. {Prosovetskii}, A.~M. {Uralov}, and
  I.~A. {Bakunina}.
\newblock {An experience of radioheliographic prediction of powerful solar
  flares}.
\newblock {\em Bulletin Crimean Astrophysical Observatory}, 106(1):31--33, June
  2010.

\bibitem{Somov1997}
Boris~V. {Somov} and Takeo {Kosugi}.
\newblock {Collisionless Reconnection and High-Energy Particle Acceleration in
  Solar Flares}.
\newblock {\em \apj}, 485(2):859--868, August 1997.

\bibitem{Su2013}
Y.~{Su}, A.~M. {Veronig}, G.~D. {Holman}, B.~R. {Dennis}, T.~{Wang},
  M.~{Temmer}, and W.~{Gan}.
\newblock {Imaging coronal magnetic-field reconnection in a solar flare}.
\newblock {\em Nature Physics}, 9(8):489--493, August 2013.

\bibitem{Su2011}
Yingna {Su}, Vincent {Surges}, Adriaan {van Ballegooijen}, Edward {DeLuca}, and
  Leon {Golub}.
\newblock {Observations and Magnetic Field Modeling of the Flare/coronal Mass
  Ejection Event on 2010 April 8}.
\newblock {\em \apj}, 734(1):53, June 2011.

\bibitem{TanakaEnome1975}
H.~{Tanaka} and S.~{Enome}.
\newblock {The Microwave Structure of Coronal Condensations and Its Relation to
  Proton Flares}.
\newblock {\em \solphys}, 40(1):123--131, January 1975.

\bibitem{Toriumi2019}
Shin {Toriumi} and Haimin {Wang}.
\newblock {Flare-productive active regions}.
\newblock {\em Living Reviews in Solar Physics}, 16(1):3, December 2019.

\bibitem{Tsuneta1997}
Saku {Tsuneta}.
\newblock {Moving Plasmoid and Formation of the Neutral Sheet in a Solar
  Flare}.
\newblock {\em \apj}, 483(1):507--514, July 1997.

\bibitem{Uralov2008}
A.~M. {Uralov}, V.~V. {Grechnev}, G.~V. {Rudenko}, I.~G. {Rudenko}, and
  H.~{Nakajima}.
\newblock {Microwave Neutral Line Associated Source and a Current Sheet}.
\newblock {\em \solphys}, 249(2):315--335, June 2008.

\bibitem{Uralov2006}
Arkadiy~M. {Uralov}, George~V. {Rudenko}, and Ilia~G. {Rudenko}.
\newblock {17GHz Neutral Line Associated Sources: Birth, Motion, and Projection
  Effect}.
\newblock {\em \pasj}, 58:21--28, February 2006.

\bibitem{vanTend1978}
W.~{van Tend} and M.~{Kuperus}.
\newblock {The development of coronal electric current systems in active
  regions and their relation to filaments and flares.}
\newblock {\em \solphys}, 59(1):115--127, September 1978.

\bibitem{Wang2017}
Haimin {Wang}, Chang {Liu}, Kwangsu {Ahn}, Yan {Xu}, Ju~{Jing}, Na~{Deng},
  Nengyi {Huang}, Rui {Liu}, Kanya {Kusano}, Gregory~D. {Fleishman}, Dale~E.
  {Gary}, and Wenda {Cao}.
\newblock {High-resolution observations of flare precursors in the low solar
  atmosphere}.
\newblock {\em Nature Astronomy}, 1:0085, March 2017.

\bibitem{Woods2017}
M.~M. {Woods}, L.~K. {Harra}, S.~A. {Matthews}, D.~H. {Mackay}, S.~{Dacie}, and
  D.~M. {Long}.
\newblock {Observations and Modelling of the Pre-flare Period of the 29 March
  2014 X1 Flare}.
\newblock {\em \solphys}, 292(2):38, February 2017.

\bibitem{Zhou2016}
G.~P. {Zhou}, J.~{Zhang}, and J.~X. {Wang}.
\newblock {Observations of Magnetic Flux-rope Oscillation during the Precursor
  Phase of a Solar Eruption}.
\newblock {\em \apjl}, 823(1):L19, May 2016.

\bibitem{Zuccarello2018}
F.~P. {Zuccarello}, E.~{Pariat}, G.~{Valori}, and L.~{Linan}.
\newblock {Threshold of Non-potential Magnetic Helicity Ratios at the Onset of
  Solar Eruptions}.
\newblock {\em \apj}, 863(1):41, August 2018.

\end{thebibliography}

\end{document}